\documentclass[preprint,1p,times]{elsarticle}

 \usepackage{graphicx, xcolor}
\usepackage{hyperref}
\usepackage{amssymb,amsmath}

\usepackage{bm}
\usepackage{siunitx}
\usepackage{booktabs}

\newcounter{bla}

\journal{Computer Physics Communications}

\begin{document}
\parindent0pt

\begin{frontmatter}



\title{ExaHyPE: An Engine for Parallel Dynamically Adaptive Simulations of
Wave Problems}

\author[a]{Anne Reinarz\corref{author}}
\author[b]{Dominic E.~Charrier}
\author[a]{Michael Bader}
\author[d]{Luke Bovard}
\author[c]{Michael Dumbser}
\author[e,g]{Kenneth Duru}
\author[f,c]{Francesco Fambri}
\author[g]{Alice-Agnes Gabriel}
\author[a]{Jean-Matthieu Gallard}
\author[d]{Sven K\"oppel}
\author[a]{Lukas Krenz} 
\author[a]{Leonhard Rannabauer}
\author[d]{Luciano Rezzolla}
\author[a]{Philipp Samfass}
\author[c]{Maurizio Tavelli} 
\author[b]{Tobias Weinzierl} 

\cortext[author] {Corresponding author.\\\textit{E-mail address:} reinarz@in.tum.de}
\address[a]{Department of Informatics, Technical University of Munich, Boltzmannstr. 3, 85748 Garching, Germany}
\address[b]{Department of Computer Science, Durham University, South Road, Durham DH1 3LE, UK}
\address[c]{Laboratory of Applied Mathematics, University of Trento, Via Messiano 77, I-38123 Trento, Italy}
\address[d]{Institute for Theoretical Physics, Goethe University, Max-von-Laue-Str. 1,  60438 Frankfurt am Main, Germany}
\address[e]{Mathematical Sciences Institute, Australian National University Canberra, Australia}
\address[f]{Max-Planck-Institute for Plasma Physics, Boltzmannstr. 2, 85748 Garching, Germany}
\address[g]{Department of Earth and Environmental Sciences, LMU Munich, Theresienstr. 41, 80333 Munich, Germany }
\begin{abstract}




ExaHyPE (``An Exascale Hyperbolic PDE Engine'') is a software engine for
solving systems of first-order hyperbolic partial differential equations (PDEs).
Hyperbolic PDEs are typically derived from the conservation laws of physics and
are useful in a wide range of application areas. Applications powered by
ExaHyPE can be run on a student's laptop, but are also able to exploit \textcolor{black}{thousands of processor cores on} state-of-the-art supercomputers.  
The engine is able to dynamically \textcolor{black}{increase the accuracy of the simulation using adaptive mesh refinement where required}. 
Due to the robustness and shock capturing abilities of ExaHyPE's numerical
methods, users of the engine can simulate linear and non-linear hyperbolic PDEs
with very high accuracy. Users can tailor the engine to their particular PDE by specifying evolved quantities, fluxes, and source terms.   
A complete simulation code for a new hyperbolic PDE can often be realised within a few hours --- a task that, traditionally, can take weeks, months, often years for researchers starting from scratch.  
In this paper, we showcase ExaHyPE's workflow and capabilities through real-world scenarios from our two main application areas: seismology and astrophysics. 
\end{abstract}


\end{frontmatter}



\noindent{\bf PROGRAM SUMMARY}\\

\begin{small}
\noindent
{\em Program Title:} ExaHyPE-Engine                                         \\
{\em Licensing provisions:} BSD 3-clause                                    \\
{\em Programming languages:} C++, Python, Fortran                                  \\


  \noindent{\em Nature of Problem:}\\
  The ExaHyPE PDE engine offers robust algorithms to solve linear and non-linear hyperbolic systems of PDEs
written in first order form. The systems may contain both conservative and non-conservative terms.  \\

{\em Solution method:}\\
 ExaHyPE employs the discontinuous Galerkin (DG) method combined 
 with explicit one-step ADER (arbitrary high-order derivative) time-stepping.
An a-posteriori limiting approach is applied to the ADER-DG solution,
whereby spurious solutions are discarded and recomputed with
a robust, patch-based finite volume scheme.
ExaHyPE uses dynamical adaptive mesh refinement to enhance the accuracy 
of the solution around shock waves, complex geometries, and interesting features.
 
\end{small}

\section{Introduction}\label{sec-introduction}

The study of waves has always been an important subject of research.
It is not difficult to see why: Earthquakes and tsunamis have a direct and serious impact on the daily lives of millions of people. 
A better understanding of electromagnetic waves has enabled ever faster wireless communication. The study of gravitational waves has allowed new insight into the composition and history of our Universe. 
Those physical phenomena, despite arising in
different fields of physics and engineering, can be modelled in a similar way from a mathematical perspective:
as a system of hyperbolic partial differential equations (PDE). 
The consortium behind
the ExaHyPE project (``An Exascale Hyperbolic PDE Engine") translated this structural similarity into  a software engine for modelling and simulating a wide range of hyperbolic PDE systems.  

\textcolor{black}{
ExaHyPE is intended as an engine, i.e. it allows only a limited number of numerical
schemes on a fixed mesh infrastructure, but provides high
flexibility in terms of the PDE system to be solved.} The consortium focuses on two challenging scenarios, long-range seismic risk assessment, see e.g. \cite{Duru:2017, tavelli:dim}; 
and the search for gravitational waves emitted by binary neutron stars, see e.g \cite{Bishop:2016,dumbser:2017, zanotti:2016, fambri:2017}.

\textcolor{black}{
A user with a given application only
needs to implement the PDE system and problem-specific well-posed initial and
boundary conditions. To exploit dynamic mesh refinement the user only needs to implement suitable criteria for mesh refinement and admissibility of solutions.
From a user's perspective, complicated PDE systems can be implemented without considering the complex issues of designing a performance-oriented high-order solver on a parallel compute cluster. Thus, the Engine enables medium-sized interdisciplinary research teams to quickly realize extreme-scale simulations of grand challenges modelled by hyperbolic conservation laws. 
}

ExaHyPE implements a high-order discontinuous Galerkin (DG) approach. DG schemes were first introduced by Reed et al. \cite{reed:1973} for the neutron transport equation and subsequently extended to general hyperbolic systems in a series of papers by Cockburn et al. \cite{shu:1,shu:2,shu:3,shu:4,shu:5}. 
Within our DG framework, higher-order accuracy in time and space is achieved 
using the \textit{Arbitrary high-order DERivative} (ADER) approach first introduced {\color{black} for purely linear constant-coefficient equations in \cite{Toro2001} and then extended to non-linear systems in  \cite{Titarev:2002}}.  
In the original ADER approach, high-order accuracy in time is achieved by using the Cauchy-Kowalevsky procedure to replace time derivatives with spatial derivatives. This procedure is very efficient for linear problems \cite{Gassner:2011:ExplicitOneStep}, but becomes cumbersome for non-linear problems. Moreover, this procedure cannot deal with stiff source terms. To tackle these shortcomings, an alternative ADER approach was introduced by Dumbser et al. \cite{Dumbser:2008}. 
In this approach, the Cauchy-Kowalevsky procedure is replaced by an implicit
{\color{black} solution} of a cell-local space-time weak formulation of the PDE. This removes the problem dependency of
the approach and allows the handling of stiff source terms.
ExaHyPE provides an implementation of both ADER-DG variants.

In non-linear hyperbolic PDE systems, discontinuities and steep gradients can
arise even from 
smooth initial conditions. 
High-order methods may then produce spurious oscillation which decrease the approximation quality and may even render the computed solution unphysical. 
To remedy this issue, a wide range of limiters for DG schemes have been proposed. 
Notably, these include approaches based on artificial viscosity \cite{hartmann:2002, persson:2006}, filtering \cite{rezzolla:2011} and WENO- or HWENO-based reconstruction \cite{shu:WENO,shu:HWENO}. 
The approach taken within the ExaHyPE engine is multi-dimensional optimal-order-detection (MOOD). 
This approach was initially applied to finite volume schemes \cite{loubere:1,loubere:2,loubere:3} and has recently been extended to DG schemes \cite{loubere:4}. 
In this approach, the solution is checked a-posteriori for certain admissibility
or plausibility criteria and is marked troubled if it does not meet them.
Troubled cells are then recalculated with a more robust finite volume scheme. 
The MOOD approach bypasses limitations of a-priori detection of troubled zones. We identify problematic regions after each time step and roll back to a more robust scheme on demand. This means that we do not need to reliably know all areas with issues a-priori. 
This approach allows for good resolution of
shocks and other discontinuities \cite{Dumbser:14:Posteriori}.

 \textcolor{black}{ The restriction to an adaptive cartesian mesh represents one of ExaHyPE's fundamental design choices.} Problems are discretised on tree-structured fully
adaptive Cartesian meshes provided by the Peano framework
\cite{Weinzierl:11:Peano,Weinzierl:2019}. 
Dynamical adaptive mesh refinement (AMR) enhances the 
shock-capturing abilities of the
ADER-DG scheme further \cite{Zanotti:2015:SpaceTimeAMR} and allows good resolution of local features.
\textcolor{black}{ExaHyPE allows for complex geometry, either handled by a curvilinear mesh transformation,  or by a diffuse interface approach. While the former approach is linear operation it is restricted to geometries that can be mapped smoothly to a cube. The latter  approach extends a given PDE system by a parameter representing the volume fraction of material in the cell and thus determines the physical boundary through a diffuse interface instead of boundary-fitted unstructured meshes. Complex geometries can thus be readily represented and extensions to moving geometries are also possible \cite{tavelli:dim,DIM2D,DIM3D}. }

ExaHyPE comprises the following key features:
\begin{itemize}
\item High-order ADER discontinuous Galerkin (ADER-DG) with a-posteriori subcell
limiting and finite volume (FV) schemes; 
\item Dynamic mesh refinement on Cartesian grids in two and three dimensions;
\item A simple API that allows users to quickly realise complex applications;
\item User-provided code can be written in Fortran or C++;
\item Automatically generated architecture- and application-specific optimised
ADER-DG routines;
\item Shared memory parallelisation through Intel’s Threading Building Blocks
(TBB);
\item Distributed memory parallelisation  with  MPI.
\end{itemize}
Furthermore, ExaHyPE offers a wide range of post-processing and plotting facilities such as support for {\color{black} the output formats vtk or tecplot}.
The software can be compiled with Intel and GNU compilers and provides a switch for choosing
different compilation modes: 
The release mode aggressively optimises the application, while 
the assertion and debug compilation modes activate the assertions within the code and print additional output. Users and developers can write log filters to filter output relevant
to them. For continuous testing a Jenkins server \cite{jenkins} verifies that the code compiles with all parallelisation features in all different compilation modes \textcolor{black}{including its ability to resolve complex geometries}.

Our paper starts with an overview of the problem setting using several examples. We then briefly sketch the solution methods used in ExaHyPE, focusing on the ADER-DG algorithm. Next we use a simple example to demonstrate the workflow when using ExaHyPE and describe the engine architecture. We conclude with a sequence of numerical examples from various application areas to demonstrate the capabilities of the engine.

\section{Problem Formulation}\label{sec-problem}
We consider hyperbolic systems of balance laws that may contain both
conservative and non-conservative terms.
They have to be given in the following first-order form:

\begin{equation}\label{eq-general-equation}
     \frac{\partial}{\partial t}  {Q} +
  \nabla\cdot  {F} ( {Q}, \nabla Q)  +
   {B}(Q)\cdot\nabla Q =  { {S}( {Q})}  + \sum_{i=1}^{n_\text{ps}} \delta_i,
\end{equation}
where $Q: \Omega\subset \mathbb{R}^d \mapsto \mathbb{R}^\nu$ is the state vector
of the $\nu$ conserved variables, $\Omega\subset\mathbb{R}^d$ is the computational domain,  $F(Q)$ is the flux tensor that may also depend on the gradient of Q in order to model viscous effects, and $B(Q)$ represents its non-conservative part.
Finally, $S(Q)$ is the source term and $\delta_i$ are the given $n_\text{ps}$ point sources.

Hyperbolic systems in the form \eqref{eq-general-equation} can be used to model a wide range of applications that involve waves. 
In the following, we demonstrate the versatility of our formulation by
introducing equations from three different application areas. Numerical experiments for these examples are provided in Section \ref{sec-numres}. 

\subsection{Waves in Elastic Media}\label{sec-elastic}
Linear elastodynamics describes waves propagating through elastic heterogeneous media by relating displacements, velocities, stress and strain. The momentum equations of motion are derived from Hooke's law and the conservation of momentum. 
Following the form of equation (\ref{eq-general-equation}) we write them as
\begin{displaymath}
 \frac{\partial}{\partial t} 
 \underbrace{
  \begin{pmatrix}
     \sigma \\ \rho v
    \end{pmatrix}}
 _{=Q}
 + \underbrace{\begin{pmatrix}
       E(\lambda,\mu) & 0\\
       0              & 0
     \end{pmatrix}
     \cdot
    \nabla
  \begin{pmatrix}
     v \\ \sigma
    \end{pmatrix}}_{=B(Q)\cdot \nabla Q}
   +
   \nabla \cdot
   \underbrace{
      \begin{pmatrix}
     0 \\ \sigma
    \end{pmatrix}
    }_{=F(Q)}
    = 0,
\end{displaymath}
where $\rho$ denotes the mass density, $v$ the velocity and the $\sigma$ stress tensor, which can be written in
terms of its six independent components as $\sigma = (\sigma_{xx}, \sigma_{yy}, \sigma_{zz},\sigma_{xy},\sigma_{xz},\sigma_{yz})$. 
In this paper we consider isotropic materials, i.e. the material matrix $E(\lambda,\mu)$ depends only on the two Lam\'e constants $\lambda$ and $\mu$ of the material.
However, the formulation holds also for more general anisotropic materials.

These equations can be used to simulate seismic waves, such as those radiated by earthquakes. 
In this context the restriction of ExaHyPE to Cartesian meshes 
seems to be restrictive. 
However, adaptive Cartesian meshes can be extended to allow the
modelling of complex topography. Two methods have been implemented in ExaHyPE
to represent complex topographies. The first approach treats ExaHyPE's adaptive Cartesian mesh as reference domain that is mapped to a complex topography
via high-order curvilinear transformations \cite{duru:curvilinear1, duru:curvilinear2}.
The second approach, a diffuse interface method, represents the topography as a smooth field \cite{tavelli:dim}. These approaches are described in more detail in Section \ref{sec-num-elast}.

\subsection{Shallow Water Equations}
In atmospheric and oceanic modelling of coastal areas, horizontal length scales are typically significantly greater than the vertical length scale.
In this case, fluid flow can be modelled with the two-dimensional 
shallow water equations instead of the more 
complicated three-dimensional Navier-Stokes equations.


Following the form of equation (\ref{eq-general-equation}) they can be written as
\begin{equation}\label{eq-swe}
\frac{\partial}{\partial t} 
\underbrace{
\begin{pmatrix}
h\\hu\\hv\\ b
\end{pmatrix}}_{=Q} + \nabla \cdot
\underbrace{\begin{pmatrix}
hu   &   hv\\
hu^2 & huv\\
huv & hv^2 \\
0 & 0\\
\end{pmatrix}}_{=F(Q)}+\underbrace{
\begin{pmatrix}
0\\
hg \, \partial_x (b+h)\\
hg \, \partial_y (b+h)\\
0\\
\end{pmatrix}}_{=B(Q)\cdot \nabla Q}= 0,
\end{equation}
where $h$ denotes the height of the water column, $(u,v)$ the horizontal flow velocity, $g$ the gravity and $b$ the bathymetry.

Hyperbolic systems of balance laws have non-trivial equilibrium solutions in which flux and source terms cancel. 
A well-balanced numerical scheme is capable of maintaining such an equilibrium state. 
In Section \ref{sec-solvers} we describe the ADER-DG scheme used in ExaHyPE and in Section \ref{sec-numres} we will describe how to keep the scheme well balanced while allowing wetting and drying.

\subsection{Compressible Navier-Stokes Equations}
ExaHyPE is extensible to non-hyperbolic equations. As an example of such an extension we show the compressible Navier-Stokes equations. They are used to model the dynamics of a viscous fluid, and are given by
\begin{equation}\label{eq:navier-stokes}
\frac{\partial}{\partial t} 
\underbrace{
\begin{pmatrix}
  \rho\\
  \rho v \\
  \rho E
\end{pmatrix}}_{=Q} + \nabla \cdot
\underbrace{\begin{pmatrix}
\rho v\\
v \otimes \rho v + I p + \sigma (Q, \nabla Q)\\
v \cdot \left( I \rho E + I p + \sigma(Q, \nabla Q) \right)
 - \kappa \nabla (T)
\end{pmatrix}}_{=F(Q, \nabla Q)}\underbrace{
=
\begin{pmatrix}
0\\
-g k \rho\\
0
\end{pmatrix}}_{=S(Q)},
\end{equation}
where $\rho$ denotes the density, $\rho v$ the momentum, $\rho E$ the energy density, $T$ the temperature and $p$ the pressure (this term includes gravitational effects). The temperature diffusion is given by $\kappa \nabla T$ with constant $\kappa$, these effects depend on the gradient of $Q$.
In the source term, the vector $k$ is the unit vector in $z$-direction and $g$ is the gravitation of Earth.
The viscous effects are modelled by the stress tensor $\sigma (Q, \nabla Q)$. Thus, the flux from \eqref{eq-general-equation} has been modified to allow $\nabla Q$ as input. 
More details on the implementation of these equations can be found in~\cite{Krenz:19:ExaCloud}.

\subsection{General Relativistic Magneto-Hydrodynamics}
The equations of classical magnetohydrodynamics (MHD) are used to model
the dynamics of an electrically ideally conducting fluid with comparable
hydrodynamic and electromagnetic forces. When modelling astrophysical objects with strong gravitational fields, e.g. neutron stars, it becomes necessary to model the background space-time as well. We use the standard $3+1$ split to decompose the four dimensional space-time manifold into 3D hyper-surfaces parameterised by a time coordinate $t$. The background space-time is introduced into the equations in the form of a non-conservative product.

Following the form of equation (\ref{eq-general-equation}) they can {\color{black} be} written as
\begin{displaymath}
\frac{\partial}{\partial t} 
\underbrace{
 \begin{pmatrix}
      \sqrt{\gamma} D\\
      \sqrt{\gamma} S_j\\
      \sqrt{\gamma} \tau\\
      \sqrt{\gamma} B^j\\
      \phi\\
      \alpha_j\\
      \beta\\
      \gamma_m
     \end{pmatrix}}_Q
     +  \nabla \cdot \underbrace{
     \begin{pmatrix}
         \alpha v^i D - \beta^i D\\
         \alpha T^i_j -\beta^iS_j\\
         \alpha(S^i-v^i D)-\beta^i\tau\\
         (\alpha v^i-\beta^i)B^j-(\alpha v^j-\beta^j)B^i\\
         0\\0\\0\\0
        \end{pmatrix}}_{=F(Q)}
+ \underbrace{
\begin{pmatrix}
 0\\
 \sqrt{\gamma} (\tau \partial_j \alpha -\frac{1}{2}T^{ik}\partial_j\gamma_{ik}-T^j_i\partial_j\beta^i) \\
  \sqrt{\gamma}(S^j\partial_j\alpha-\frac{1}{2}T^{ik}\beta^j\partial_j\gamma_{ik}-T^j_i\partial_j\beta^j)\\
  -\beta^j\partial_i(\sqrt{\gamma}B^i)+ \alpha\sqrt{\gamma} \gamma^{ji}\partial_i \phi\\
   \sqrt{\gamma}\alpha c_h^2\partial_j(\sqrt{\gamma}B^i)-\beta^j\partial^j\phi\\
   0\\0\\0
\end{pmatrix}}_{=B(Q)\cdot \nabla Q}
= 0,
\end{displaymath}
where $i,j=1,2,3$ and $m=1...6$.

The curved space-time is parameterised by several hyper-surface variables:
lapse $\alpha$, spatial metric tensor $\gamma$ shift vector $\beta$
and extrinsic curvature $K$. The spatial metric tensor is given as a vector of its six independent components $\gamma=(\gamma_{11},\gamma_{12}, \gamma_{13},\gamma_{22},\gamma_{23},\gamma_{33})$  and has the determinant $\sqrt{\gamma}:=\det{\gamma}$.
Further, $D=W\rho$ is the conserved density, which is related to the rest mass density $\rho$ by the Lorentz factor $W$, $v^i$ is the fluid velocity, $T$ is the Maxwell 3-energy momentum tensor, $S$ is the conserved momentum, $B$ is the magnetic field and $\tau$ is the conserved energy density. Finally, $\phi$ is an artificial scalar introduced to ensure a divergence-free magnetic field, and $c_h$ is the characteristic velocity of the divergence cleaning. 
For more details on this formulation see e.g. \cite{koeppel:grmhd,fambri:grmhd,rezzolla:2013}.

\section{Solver Components}\label{sec-solvers}
\textcolor{black}{
In ExaHyPE's engine concept the numerical method is given and in general the user does not need to interact with any solver components. A layer of generated glue code separates the user from the kernel calls. The main kernels are given in Table \ref{tbl:kernels}, of these only the slope-limiter and Riemann solver can be modified by the user. However, all modifications to the kernels are optional since stable defaults are provided. This section briefly summarises the numerical algorithms used. } 
 For brevity, let us use a pared down form of (\ref{eq-general-equation}), in which only the flux term is nontrivial:
\begin{equation}
\frac{\partial Q}{\partial t} 
+
\nabla \cdot F(Q)
=0
\qquad\rm{on}\;\Omega \subset \mathbb{R}^d,\; d=2,3.
\end{equation}
Assume that \eqref{eq-general-equation} is subject to appropriate initial and boundary conditions:
\begin{equation*}
 Q(x,0)=Q_0(x),~\forall x\in \Omega,~~~~~Q(x,t)=Q_B(x,t),~\forall x\in \partial\Omega,\,\forall t\in\mathbb{R}^+_0.
\end{equation*}

\begin{table}
     \centering
    \caption{An overview of the main kernels implementing the algorithm, all are implemented for 2D and 3D and in linear and non-linear versions.}
    \label{tbl:kernels}
\textcolor{black}{
    \begin{tabular}{ll}
       \toprule
        \multicolumn{2}{c}{\bf ADER-DG}\\
        \midrule
        spaceTimePredictor & Linear: Cauchy-Kowalevsky,\\
                           & Nonlinear: weak element-local space-time DG\\
        Integrals & Implements face, surface, and volume integration\\
        \midrule
        \multicolumn{2}{c}{\bf FV}\\
        \midrule
        Godunov       & Implements the Godunov FV scheme\\
        MUSCL-Hancock & Implements the MUSCL-Hancock FV scheme\\
        slopeLimiter  & Various slope limiters , default: minmod\\
        \midrule
        \multicolumn{2}{c}{\bf Both (ADER-DG + FV)}\\
        \midrule
        solutionUpdate     & Updates the solution vector\\
        riemannSolver      & Riemann solver, default: Rusanov\\
        stableTimeStepSize & Calculates next stable time step size\\
        \midrule
        \multicolumn{2}{c}{\bf Limiter}\\
        \midrule
        projectOnFVLimiterSpace  & Projects DG solution on FV subcells\\
        projectOnDGSpace         & Projects FV solution back onto DG grid\\
        discreteMaximumPrinciple & Limiter criterion, checks discrete maximum principle\\
       \bottomrule
    \end{tabular}}
\end{table}

\subsection{Discretisation}
\begin{figure}[tb]
\centering
 \includegraphics[width=0.5\textwidth]{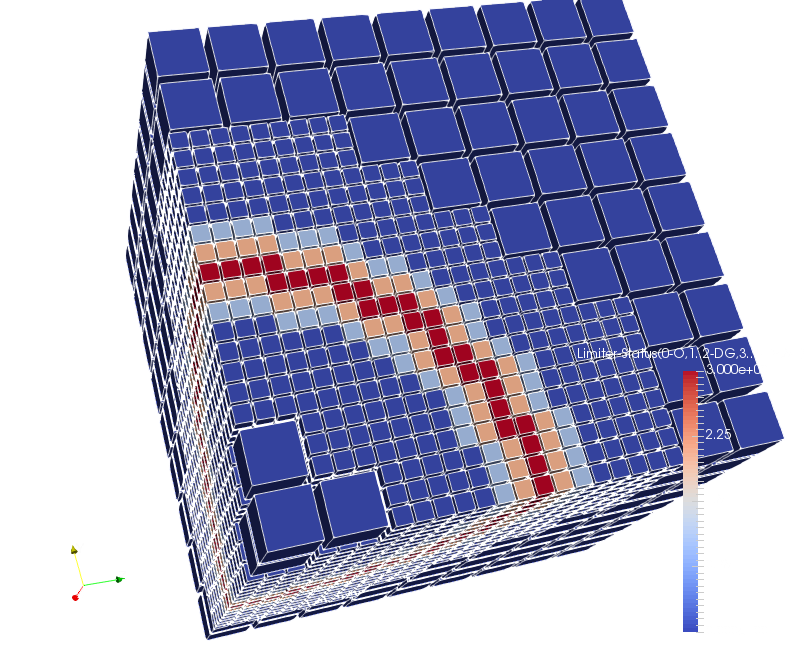}
 \hfill
 \includegraphics[width=0.4\textwidth]{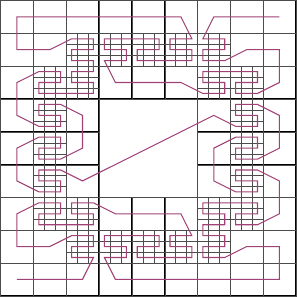}
 \caption{Left: Adaptive Cartesian mesh in 3 dimensions. The FV limiter is active on the finest level. Right: 
 Peano space-filling curve running through a 2D mesh that has a similar refinement pattern.
  }
 \label{fig:amr}
\end{figure}

ExaHyPE allows for Cartesian grids in two and three dimensions. 
The computational domain is divided into a grid $\Omega = \bigcup_i T_i$ using a
space-tree construction scheme \cite{Weinzierl:11:Peano,Weinzierl:18:Peano}.
Each cell $T_i$ is recursively refined giving an adaptive Cartesian grid as shown
in Figure \ref{fig:amr}. Meshes can be defined with an arbitrary numbers of elements in each direction on the coarsest level. Refinement is based on tripartitioning.

{\color{black}
The rationale behind a commitment to tripartitioning is two-fold.
On the one hand, we rely on the Peano AMR framework \cite{Weinzierl:18:Peano}
which internally orders all cells along a Peano space-filling curve.
This yields a high temporal and spatial locality of the cell and face accesses;
a property the Peano curve shares with other space-filling curves.
However, the Peano curve also allows us to realise all temporary data structures
through cache-friendly stacks \cite{Weinzierl:11:Peano}.
On the other hand, tripartitioning is an intrinsic match for patch-based Finite
Volume methods which work with odd numbers of volumes per axis:
A subdivision of a cell here makes a former volume centre coincide with a volume
centre on the next finer level.
This simplifies the structure of inter-resolution transfer operators and
preserves them over many resolution levels.
}

\subsection{Finite Volumes}\label{sec:fv} 
Classically systems of hyperbolic PDE have been solved using finite difference or finite volume schemes. In a finite volume method cell averages are calculated. Volume integrals in a PDE that contain a divergence term are converted to surface integrals using the divergence theorem. These terms are then evaluated as fluxes at the surfaces of each element. However, to achieve high order accuracy in a finite volume scheme, large stencils and expensive recovery or reconstruction procedures are needed. Examples include essentially non-oscillatory (ENO) or weighted ENO (WENO) schemes, see e.g. \cite{abgrall:1994, dumbser:weno}.

\textcolor{black}{As shown in Table \ref{tbl:kernels}} ExaHyPE provides access to classical Godunov type schemes as well as MUSCL schemes \cite{muscl, leer:1997}. Further, users can use parts of the generic compute routines while implementing other parts on their own. For instance, in both the ADER-DG and Finite Volume schemes, users can overwrite the Riemann Solver with their own implementation. 

\subsection{ADER-DG} 

 \textcolor{black}{Instead of representing the solution as cell averages, DG methods represent the solution  within each cell by a (high-order) polynomial. The ADER-DG method is a one-step predictor-corrector scheme. The  integration in time is initially performed only within the element, neglecting the element interfaces and then a single correction step is performed to take the element interfaces into account \cite{ader1}.}

 The method operates on a weak form of equation (\ref{eq-general-equation}) \begin{align}\label{eq-solvers-prediction}
\int_{T_i}\int_{t^n}^{t^{n+1}} 
\theta_h\,\frac{\partial q_h}{\partial t}\,
\text{d}x\text{d}t
+
\int_{T_i}\int_{t^n}^{t^{n+1}}
\theta_h\,\nabla\cdot F(q_h)\,
\text{d}x\text{d}t
= 0.
\end{align}
{\color{black}This formulation has replaced the solution $Q$ with a discrete function $q_h$, represented by Lagrange polynomials $\theta_h$ in time and space. 
This solution is refered to as the space-time predictor and intuitively corresponds to a Taylor-expansion in space and time. 
Hence, it is equal to the time-evolution of the discrete solution within the control element for a smooth solution.
The space-time test function is in the space of piecewise polynomials, which is constructed as tensor products of Lagrange polynomials over Gauss-Legendre or Gauss-Lobatto points.

Integrating over each element $T_i$ and over the current time interval $[t^n,t^{n+1}]$ then gives the element-local weak formulation.
 Note that the formulation does not take into account any information from neighboring elements. The predictor is subsequently corrected using contributions from neighboring cells using a Riemann-solver. 
 
 However, first we need to compute $q_h$.
 This can be evaluated directly by the Cauchy-Kovalevsky algorithm for linear problems.
 For non-linear models, we use the ADER-DG method proposed by Dumbser et al. \cite{Zanotti:2015}.
It is a fixed-point iteration and can be seen as a discrete counterpoint to the well-known Picard iteration, see \cite{Dumbser:14:Posteriori}.
}
There are three phases per ADER-DG time step:
\begin{enumerate}
 \item \textcolor{black}{Per grid cell $T_i$ and time interval $[t^n,t^{n+1}]$, we first implicitly solve (\ref{eq-solvers-prediction}), predicting the local evolution. The concurrent solves of (\ref{eq-solvers-prediction}) do not take into account any information from neighbouring elements and thus yield jumps along the cell faces in $q_h$ and $F(q_h)$.}
\item \textcolor{black}{We traverse all faces of the grid and compute a numerical normal flux $G(q_h, F(q_h))$ from both adjacent cells.}
ExaHyPE uses a Rusanov flux by default; users can replace it with any other Riemann solver. 
\item In the corrector step, we traverse the cells again and solve
\begin{align}
\label{eq-solvers-correction}
\int_{T_i} \theta_h\,\Delta q_h\,\text{d}x
=
-\int_{T_i}\int_{t^n}^{t^{n+1}} 
\nabla \theta_h \cdot F(q_h)\,\text{d}x\text{d}t 
+\int_{\partial T_i} \int_{t^n}^{t^{n+1}}\theta_h\,G(q_h,F(q_h))\,\text{d}s\text{d}t
\end{align}
for $\Delta q_h = q_h(t^{n+1}) - q_h(t^n)$, see \cite{Dumbser:2008}.
The one-step update (\ref{eq-solvers-correction}) is \textcolor{black}{ derived by multiplying with a spatial testfunction} and partially integrating (\ref{eq-general-equation}).
Equation \ref{eq-solvers-correction} can be easily inverted given that the
ansatz and test space typically yield a diagonal mass matrix. 
\end{enumerate}

\subsection{Time-Step Restrictions}
Nonlinear effects and mesh adaptation require
adjustments to the time step size during a simulation.
This is expressed by the CFL condition, which
gives an upper bound on the stable time step
size for explicit DG schemes:
\begin{equation}\label{eq-cfl}
  \Delta t \leq \frac{\text{CFL}_N}{d\,(2N+1)}\,\frac{h}{|\lambda_\text{max}|},
\end{equation}
where $h$ and $|\lambda_\text{max}|$ are the mesh size and the {\color{black} maximum signal velocity}, respectively,
and $\text{CFL}_N < 1$ is a stability factor that depends 
on the polynomial order \cite{Dumbser:2008}.

 \subsection{A-Posteriori Limiting}
The unlimited ADER-DG algorithm will suffer from numerical oscillations (Gibbs phenomenon) in the presence of steep gradients or shock waves. Therefore a limiter must be applied.
 The approach followed in ExaHyPE is based on the a-posteriori MOOD method of Loub\`ere et al. \cite{loubere:4}. The solution is checked a-posteriori for certain admissibility
or plausibility 
 criteria and is recalculated with a robust FV scheme if it does not meet them.  The FV patch size is chosen to have an order of $2N+1$, this is the smallest cell size that does not violate the CFL condition \eqref{eq-cfl}. 
 
\begin{figure}[t]
  \centering
  \includegraphics[width=0.4\textwidth]{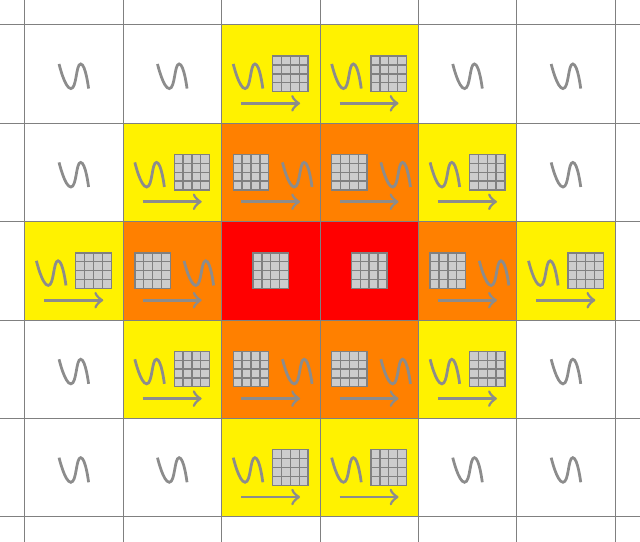}
  \hspace{0.02\textwidth}
  \includegraphics[width=0.4\textwidth]{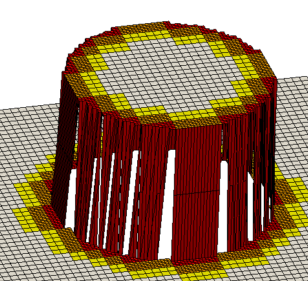}
 \caption{
  Left: Limiter status stencil. Two FV cells (red) are surrounded by (ADER-)DG cells
  (white). Cells in the interface layers compute with FV and project to DG (orange)
  or with DG and project to FV (yellow).
  Right: The FV and DG subdomains are initialised along a discontinuity.
  }
  \label{fig-hybrid-limiter}
\end{figure}

 In contrast to the original approach, ExaHyPE's approach incorporates the observation that cells usually require a recalculation with FV multiple time steps in a row after the initial check failed. Therefore, ExaHyPE implements the a-posteriori limiting ADER-DG method as a hybrid ADER-DG-FV method (Figure \ref{fig-hybrid-limiter}). 

\goodbreak

As a-posteriori detection criteria we use
\begin{enumerate}
 \item \textit{Physical Admissibility Criteria}: Depending on the system of PDEs being studied certain physical constraints can be placed on the solution. For the shallow water equations these are positivity of the water height. These criteria are supplied by the user along with the PDE terms.
 \item \textit{Numerical Admissibility Criteria}: To identify shocks we use a relaxed discrete maximum principle (DMP) in the sense of polynomials, for details see e.g. \cite{loubere:4}. 
 \end{enumerate}

We call DG cells that do not satisfy the above criteria 
\textit{troubled} cells.
If a cell is flagged as troubled and has not been troubled in the previous time step, the scheme goes back to the old time step 
 and recomputes the solution in all troubled cells (and their direct neighbours) with the FV scheme.

\section{Engine Architecture and Programming Workflow}\label{sec-arch}

\begin{figure}[t]
\centering
 \includegraphics[width=0.7\textwidth]{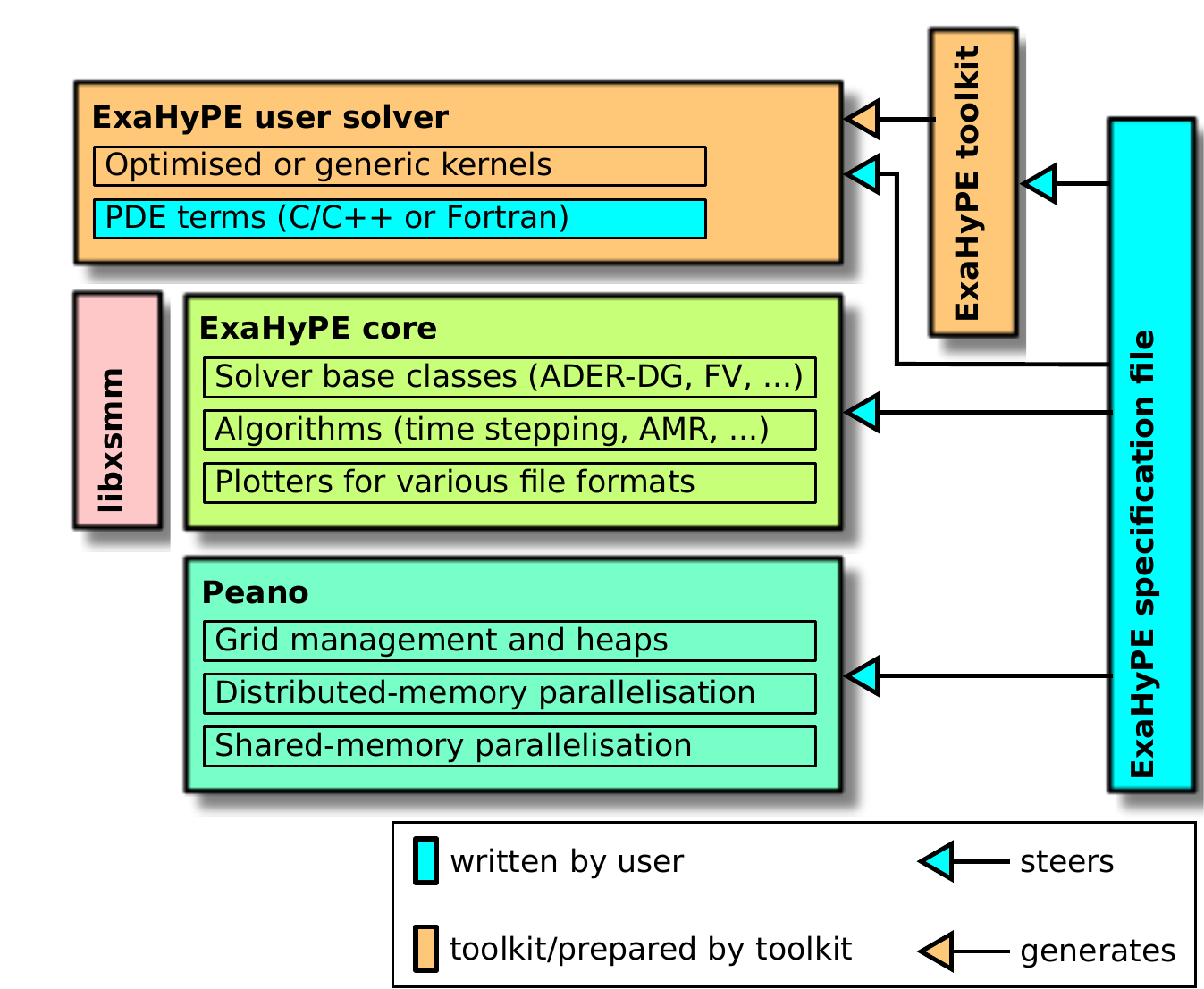}
 \caption{Engine architecture}
 \label{fig:arch}
\end{figure}

 This section briefly walks through the workflow of setting up an application in ExaHyPE using the shallow water equations given in \eqref{eq-swe} as an example. 
The architecture of ExaHyPE is illustrated in Figure \ref{fig:arch}. ExaHyPE is a solver engine,  domain-specific code has to be written by the user to obtain simulation code. In Figure \ref{fig:arch} a turquoise colour is used to highlight files written by the user. To write an ExaHyPE application users typically start from a specification file. The specification file is passed to the ExaHyPE toolkit, which creates glue code, empty application-specific classes and optionally application and architecture tailored core routines. The application specific classes are filled by the user with the PDE terms. This code can be written in C++ or Fortran. The generated glue code and the initially empty templates make up the ExaHyPE user solver.

{\color{black} The ExaHyPE core relies on the Peano framework (green) for its
dynamically adaptive Cartesian meshes.} In addition, it provides
an efficient mesh traversal loop that ExaHyPE's algorithms plug into. Peano itself is a third-party component. 

The number of dependencies in the ExaHyPE core is minimal.  However, the architecture may not fulfil the requirements of all applications, so the user can extend and connect further software fragments to the ExaHyPE core. In red we show an optional dependency, libxsmm \cite{Heinecke:16:LIBXSMM}. This package provides efficient kernels for small matrix multiplications, which are used by the optimised ADER-DG routines described in Section \ref{sec-optker}. Similarly, further software packages can be added by the user as needed.

\textcolor{black}{In Table \ref{tbl:usrfunctions},  a brief summary of the components of the user solver is given. In the following Section we highlight the parts of the specification file that need to be modified for each component and show the implementation of the user functions for the shallow water equations.} 
The solver components can be set up flexibly by modifying the specification file. Users only need to write application-specific code that sets up their PDE system.

\begin{table}[bt]
     \centering
    \caption{An overview over the optional and required user-implemented functions.}
    \label{tbl:usrfunctions}
\textcolor{black}{
    \begin{tabular}{ll}
       \toprule
        \multicolumn{2}{c}{\bf Required Functions}\\
       \midrule
        adjustPointSolution & Set values in the state vector. \\
                            & Required: initial values. Optional: constraints\\
        eigenvalues & Eigenvalues to calculate time step restriction\\
        boundaryValues & Define boundary conditions\\
        \midrule
        \multicolumn{2}{c}{\bf Problem-dependent Functions}\\
        \midrule
        flux & flux function\\
        nonConservativeProduct  & non-conservative products\\
        source, pointSource & source terms\\
        multiplyMaterialParameterMatrix & material parameters\\
        viscousFlux & Flux terms with access to gradients\\
        viscousEigenvalues & Eigevalues for time step restriction in viscous terms\\
        \midrule
        \multicolumn{2}{c}{\bf Optional Functions}\\
        \midrule
        init                & Initialisation of external libraries \\
        refinementCriterion & Only for applications with AMR\\
                            & define additional areas to refine\\
        isPhysicallyAdmissable & Only for applications with Limiter\\
                               & Gives physical admissibility criteria\\
       \bottomrule
    \end{tabular}}
\end{table}

\subsection{Code Generation and Compilation}
After the specification file has been written it is handed over to the ExaHyPE toolkit. A minimal ExaHyPE specification file is shown below.
\begin{verbatim}
 exahype-project SWE
  output-directory const = ./SWE

  computational-domain
    dimension const         = 2
    width                   = 1.0, 1.0
    offset                  = 0.0, 0.0
    end-time                = 1.0
  end computational-domain
  
  solver ADER-DG MySWESolver
    variables const    = h:1,hu:1,hv:1,b:1
    order const        = 3
    maximum-mesh-size  = 0.1
    time-stepping      = global
    type const         = non-linear
    terms const        = flux,ncp
    optimisation const = generic, usestack
  end solver
end exahype-project
\end{verbatim}

To prepare this example for the simulation, run
\begin{verbatim}
> ./Toolkit/toolkit.sh Demonstrators/SWE.exahype
\end{verbatim}

The toolkit generates a Makefile, glue code, as well as various helper files. Among them is one C++ class per solver that was specified in the specification file. Within each implementation file, the user can specify
initial conditions, mesh refinement control, etc. For example, to set the eigenvalues the following function is generated in the file  \verb=MySWESolver.cpp=.
\begin{verbatim}
void SWE::MySWESolver::eigenvalues(const double* const Q,
                                   const int direction,
                                   double* const lambda) {
  // @todo Please implement/augment if required
  lambda[0] = 1.0;
  lambda[1] = 1.0;
  lambda[2] = 1.0;
  lambda[3] = 1.0;
}
\end{verbatim}
This function can then be filled with the eigenvalues of the PDE system under consideration. Similarly, the other functions flux, initial conditions and boundary conditions can be defined in the file. For the shallow water equations this would be:
\begin{verbatim}
void SWE::MySWESolver::eigenvalues(const double* const Q,
                                   const int direction,
                                   double* const lambda) {
  ReadOnlyVariables vars(Q);
  Variables eigs(lambda);

  const double c = std::sqrt(gravity*vars.h());
  double u_n = Q[direction + 1] * 1.0/vars.h();

  eigs.h()  = u_n + c;
  eigs.hu() = u_n - c;
  eigs.hv() = u_n;
  eigs.b()  = 0.0;
}
\end{verbatim}
In this example we have used named variables to enhance readability. However, ExaHyPE also allows the user to access the vectors directly as can be seen in the generated function above.

\begin{figure}[t]
  \centering
  \includegraphics[width=0.9\textwidth]{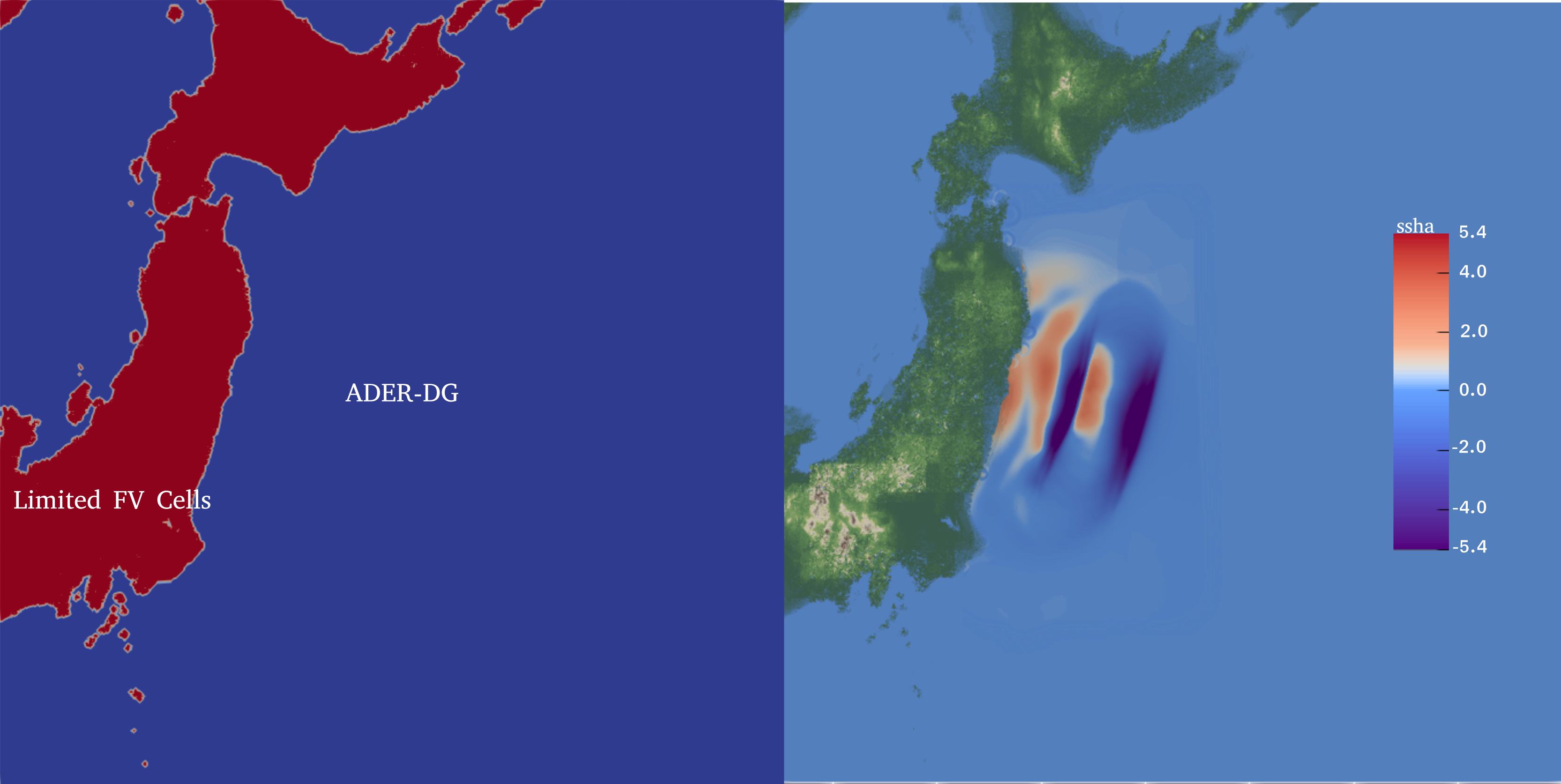}
 \caption{
  Vtk output of a shallow water simulation of the Tohoku tsunami. Left: FV and DG domains. Right: the tsunami 5 min after the initial event. These figures are taken from \cite{rannabauer:swe:2018}.
  }
  \label{fig-euler}
\end{figure}

The whole build environment is generated. A simple make will create the ExaHyPE executable. ExaHyPE's specification files always act as both specification and configuration file, i.e when running the code the specification file is passed in again.
\begin{verbatim}
> ./ExaHyPE-SWE ./SWE.exahype
\end{verbatim}

A successful run yields a sequence of \texttt{vtk} files that you can open with Paraview or VisIt. In this example, we plot two quantities. 
Such an output is shown in Figure \ref{fig-euler}.

Global metrics such as integrals can also be realised using a plotter with no output variables, i.e. \texttt{variable const = 0}.

\section{Parallelisation and Optimisation Features}
ExaHyPE relies on the Peano framework for mesh generation. Peano implements
cache-efficient, tree-structured, adaptive mesh refinement. To traverse the cells or vertices the mesh traversal automaton of Peano runs along the Peano space-filling-curve (SFC), see Figure \ref{fig:amr}.
The action of a PDE operator is mapped to the SFC traversal by 
plugging into events triggered by the traversal automaton, these mappings can be generated by a toolkit.

Peano also takes care of the distributed-memory and shared-memory parallelisation. Domain decomposition for distributed-memory
parallel simulations is realised by forking off or merging subtrees. The domain decomposition can be influenced via load balancing callbacks.
The shared-memory parallelisation relies on identifying regular substructures in the tree and employing parallel-for loops in these areas \cite{Weinzierl:RunLengthEncoding}.
Recently, we introduced a runtime tasking system to Peano
and ExaHyPE that introduces additional multi-threading concurrency \cite{Charrier:18:Enclave} in highly adaptive mesh regions and allows overlapping computations with MPI communication.


ExaHyPE builds on the Peano framework and it inherits a user model from Peano: Our engine removes the responsibility for algorithmic issues from the user. The time step, the mesh traversal and the parallelisation are implemented in a generic way. The user only controls the PDE system being solved, by specifying how many quantities are needed, which terms are required, and ultimately what all terms look like.
Our goal is to allow users to focus on the physics only and to hide away as
many implementation details as possible.

\subsection{High-Level Optimisations}

ExaHyPE contains several high level algorithmic optimisations that users can switch
on and off at code startup through the specification file. 
To gain access to these optimisations, we add the following optional section to the minimal specification file shown in Section \ref{sec-arch}.
\begin{verbatim}
global-optimisation
    fuse-algorithmic-steps               = all
    spawn-predictor-as-background-thread = on
    spawn-amr-background-threads         = on
end global-optimisation
\end{verbatim}

\noindent
A discussion of each individual algorithmic tuning is beyond scope of this paper.
Some techniques are:
\begin{itemize}
 \item \textbf{Step fusion:} Each time step of an ExaHyPE solver consists of three phases: computation of local ADER-DG
predictor (and projection of the prediction onto the cell faces), solve of the Riemann problems
at the cell faces, and update of the predicted solution. We may speed up the code if we fuse these
four steps into one grid traversal. 
\item \textbf{Spawn background jobs:} TBB has a thread pool in which idle threads wait for tasks, as soon as new tasks are spawned the threads in this pool start this task. We refer to these threads as background threads. Since spawning and scheduling these tasks has a overhead we wait until a certain number of tasks has accumulated and schedule these tasks together.
%
Certain space-time-predictor computations can be spawned as a background job. Costly AMR operations such as the imposition of initial conditions and evaluation of refinement
criteria can also be performed as a background jobs.
\item \textbf{Modify storage precision: }ExaHyPE internally can store data in less-than-IEEE double precision. This can reduce the memory
footprint of the code significantly.
\end{itemize}

Most users will not modify these options while they prototype. 
When they start  production runs, they can tweak the engine instantiation through them.
The rationale behind exposing these control values is simple: It is not clear at
the moment which option combinations robustly improve performance. 
An auto-tuning/machine learning approach to find the optimal  parameter combinations
could be considered..

ExaHyPE allows running multiple simulations on the same computational
mesh. This can be facilitated by adding multiple solver descriptions
to the specification file. Each solver uses its own base grid and 
refinement criterion.

\subsection{Optimised ADER-DG Routines}\label{sec-optker}
One of ExaHyPE's key ideas is to use tailored, extremely optimised code routines
whenever it comes to the evaluation of the ADER-DG scheme's steps and other computationally expensive routines.
AMR projections and the space-time predictor \eqref{eq-solvers-prediction} are typically the dominant steps.
If the user decides to use these optimised variants in the specification file, the toolkit calls a specific code generator Python3 module and links to its output in the generated glue code so that the calls to the generic routines are replaced by calls to the generated optimised ones.

SIMD operations, notably introduced with AVX-512 on KNL and Skylake, become
increasingly critical to fully exploit the potential of modern CPUs.
Therefore, on Intel machines, the optimised routines' main goal is to either directly use
SIMD or enable as much auto-vectorisation from the compiler as possible.
To that end the optimised routines use Intel's libxsmm \cite{Heinecke:16:LIBXSMM}, which is the second third-party building block in the basic ExaHyPE architecture.
We map all tensor operations required by the ADER-DG algorithm to general matrix multiplications (gemms) and apply architecture-specific vectorised matrix operations.
Furthermore, the code generation 
allows the introduction of data padding and alignment to get the most out of the compiler auto-vectorisation.

While improving SIMD vectorisation is our current prime use case for the optimised kernels, we
provide alternative optimisation dimensions:
Instead of a vectorisation of the native ADER-DG loops,
the optimised routines can also be configured to work with vectorised PDE
formulations.
This implies that the PDE terms are computed through SIMD operations instead of scalar ones and requires some work from the user.
This has been found to greatly improve performance \cite{Dumbser:2018}.

{\color{black} Instead of the standard Picard iteration, we provide a continuous extension Runge-Kutta scheme (CERK) to yield an initial guess to the Picard
iteration}, leading to a much lower number of required iterations.


\section{Numerical Results}\label{sec-numres}

In this section, we show the ExaHyPE engine in action. 
The applications in this section are taken from our introductory discussion in Section \ref{sec-problem}.  Only the Euler equations are added as an all-time classic starting point for solvers of hyperbolic systems.
All specification and source files used to generate the results in this section
are made publicly available, and all of the tests themselves can be
run on a standard laptop. The scaling tests shown require the use of a larger cluster.

\subsection{Euler Equations}
 The non-linear compressible Euler equations model the flow of an inviscid fluid with constant density. Solutions of the Euler equations are sometimes used as approximations to real fluids problem, e.g. the lift of a thin airfoil.
They are given by
\begin{equation}\label{eq-euler}
\frac{\partial}{\partial t} \begin{pmatrix}
\rho\\j\\\ E
\end{pmatrix}
+
\nabla\cdot\begin{pmatrix}
{j}\\
\frac{1}{\rho}j\otimes j + p I \\
\frac{1}{\rho}j\,(E + p)
\end{pmatrix}
= 0,
\end{equation}
where $\rho$ denotes the mass density, $j\in\mathbb{R}^d$ denotes the momentum density, 
$E$ denotes the energy density, $p$ denotes the fluid pressure, to be given by an equation of state, and $\otimes$ denotes the outer product.

{\color{black}
We extend the model by a color function denoting the volume fraction of material in a cell starting from the Baer-Nunziato model \cite{BaerNunziato1986}: 
\begin{equation} \label{eq-euler-dim}
\frac{\partial}{\partial t} \begin{pmatrix}
\alpha\rho \\
\alpha j\\
\alpha E\\
\alpha
\end{pmatrix}
+
\nabla\cdot\begin{pmatrix}
\alpha j\\
\frac{\alpha}{\rho}j\otimes j + \alpha p I \\
\frac{\alpha}{\rho}j\,(E + p) \\
0
\end{pmatrix}
+
\begin{pmatrix}
 0\\
 -p\nabla \alpha\\
 0\\
 0
\end{pmatrix}
= 0,
\end{equation}
In the above PDE system $\alpha$ denotes the volume fraction of material present. The pressure is given by:
\begin{equation}
 p= (\gamma-1)\left( E - \frac{1}{2\rho} \|j\|^2_2\right), 
\end{equation}
where $\gamma = 1.4$ is the ratio of specific heats .

Note that the addition of $\alpha$ has introduced a non-conservative term into the equation.
In the above we have given a \textit{reduced} Baer-Nunziato model, similar to the approach presented in \cite{DIM2D,DIM3D,Gaburro2018,tavelli:dim}. This model allows simulation of \textit{fluid-structure-interaction (FSI)} problems on adaptive Cartesian grids, without requiring boundary-fitted grids.


In Figure \ref{fig-airfoil} we present computational results for flow over a thin airfoil.
Initial conditions were:
 \begin{displaymath}
 j(x,0)=(1,0),~ \rho(x,0) = 1.0, ~ E(x,0) = 2.5, 
 \end{displaymath}
 As boundary conditions we set an inflow boundary at the left, an outflow boundary at the right and $v\cdot\nabla n=0$ at the top and bottom boundaries. Results are shown for a NACA 4612 airfoil \cite{NACA}. The results generate the expected bow shock. 
\begin{figure}[tb]
\centering
\includegraphics[width=0.48\textwidth]{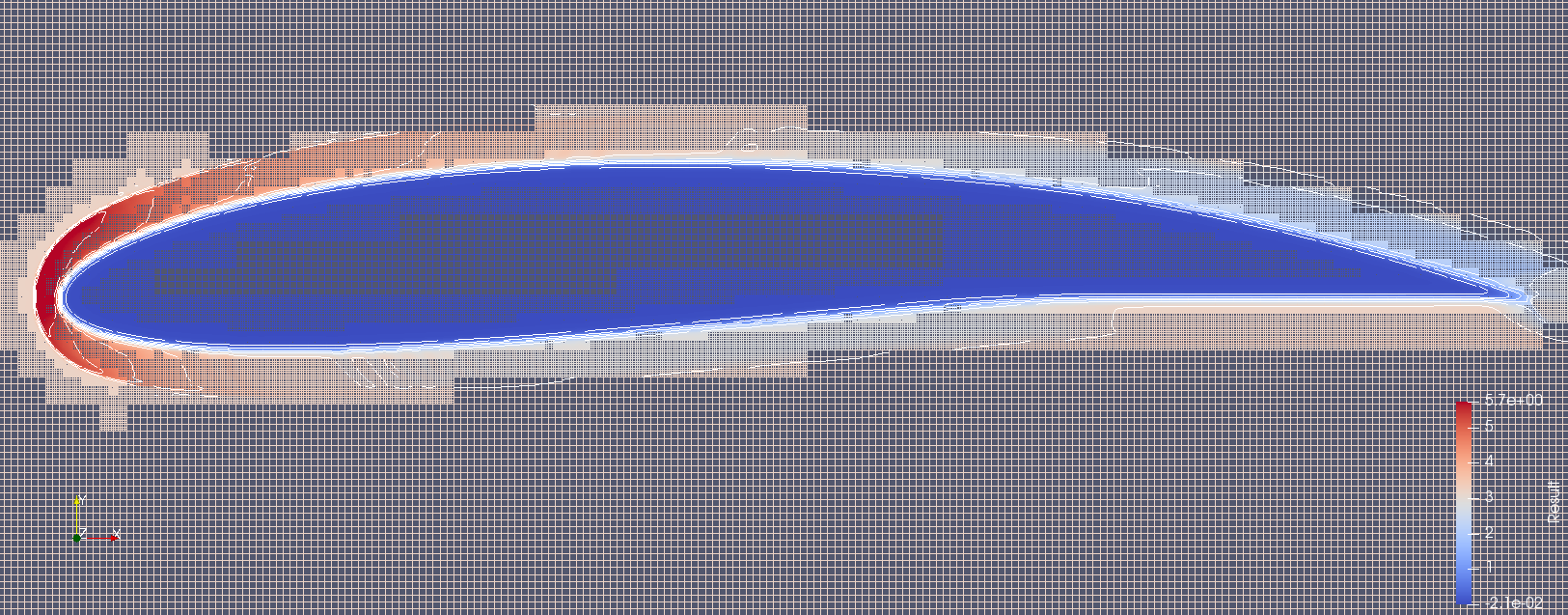}\hspace{0.5em}
\includegraphics[width=0.48\textwidth]{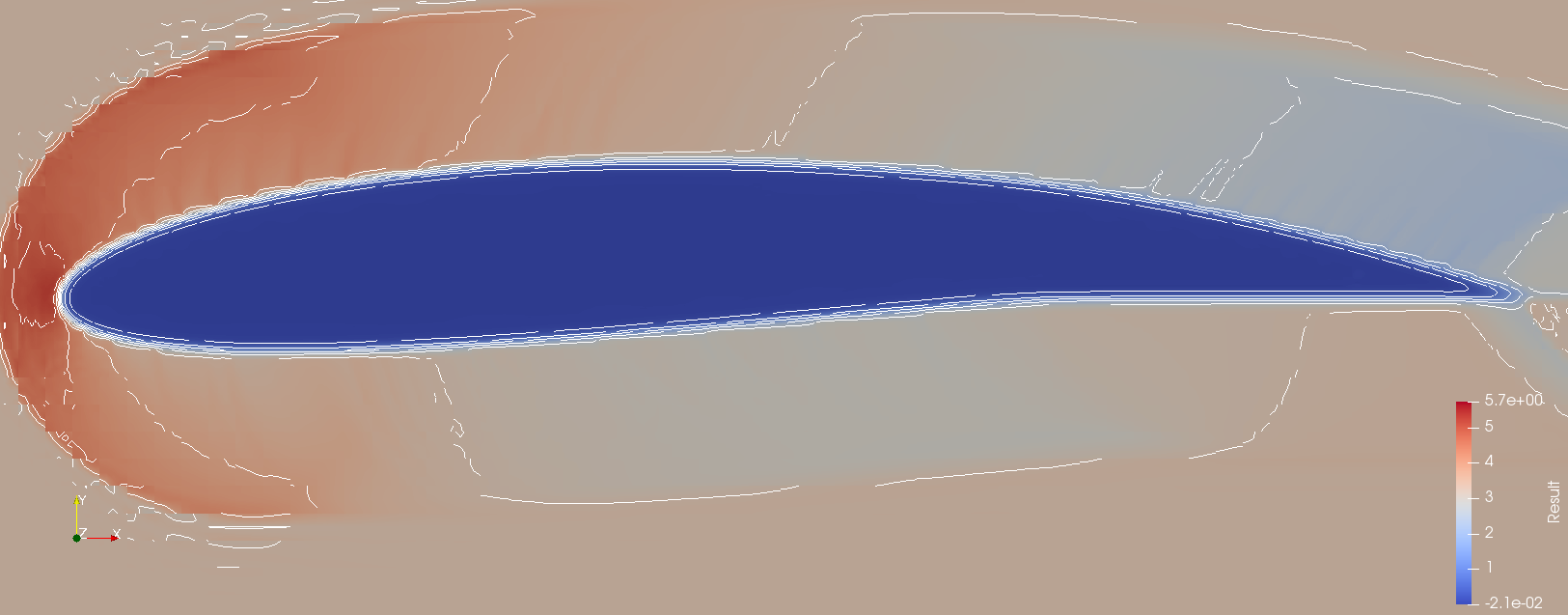}
\caption{An ADER-DG implementation of flow over an airfoil in 2D using a base grid of $79\times 25$ cells. From left to right: The energy $E$ at $t=0.03$ and at $t=0.1$. }
\label{fig-airfoil}
\end{figure}

}

\begin{figure}[h]
\centering
\includegraphics[width=0.45\textwidth]{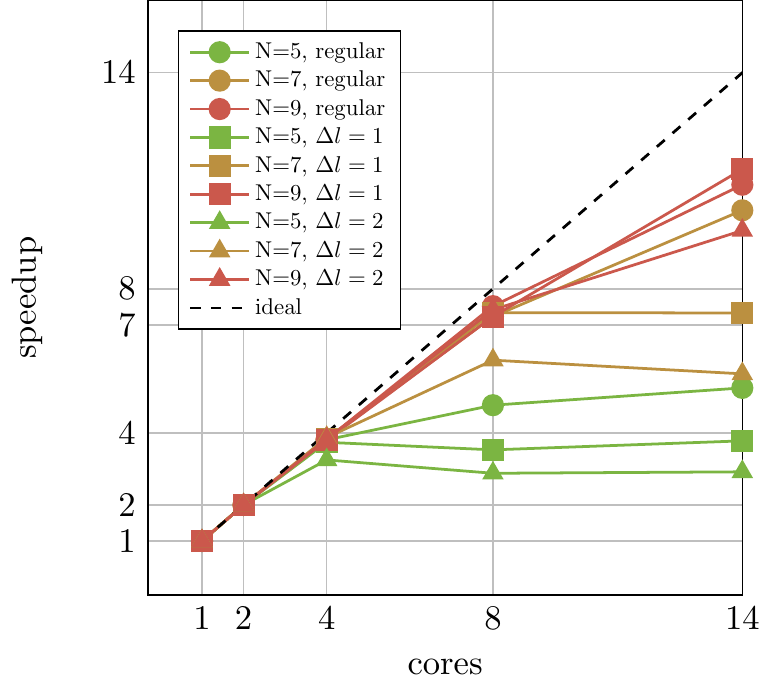}
\hspace{0.05\textwidth}
\includegraphics[width=0.45\textwidth]{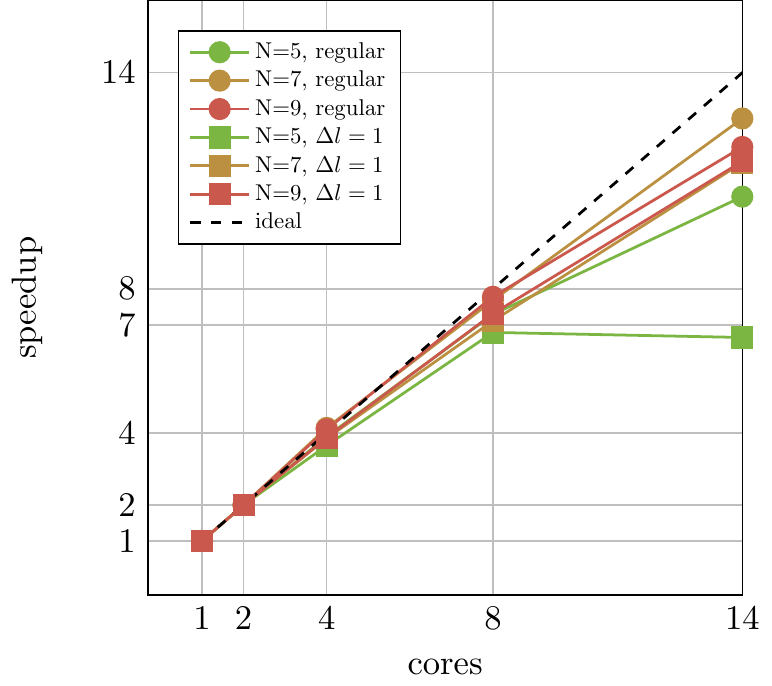}
\caption{ Shared-memory scalability of the non-linear ADER-DG implementation for a circular, spherical explosion problem  experiments using a base grid with up to two levels of adaptive refinement. Left: 2D test with a base grid of  of $81^2$ cells. Right: 3D test with a base grid of $27^3$ cells.}
\label{fig-euler-explosion-tbb}
\end{figure}

To demonstrate the shared-memory scalability of the code we use a variant of the Sod shock tube problem, the circular, spherical explosion problem, also referred to as a ``multi-dimensional Sod shock tube''. This test case uses the following initial conditions
 \begin{displaymath}
 j(x,0)=0,~ \rho(x,0) = \begin{cases}
                  1 & \text{if } \|x-x_0\|^2<r^2,\\
                  \frac{1}{8} & \text{else,}
                 \end{cases}, ~ E(x,0)=\begin{cases}
                  1 &  \text{if } \|x-x_0\|^2<r^2,\\
                  \frac{1}{10} & \text{else.}
                 \end{cases}
 \end{displaymath}
and wall boundary conditions. 

In Figure \ref{fig-euler-explosion-tbb} we show the shared-memory scalability in
two and three dimensions. This scaling test was run on the $14$ cores of an Intel
Xeon E5-v3 processor with $2.2$ GHz core frequency (SuperMUC Phase~2). 
The tests were run on one processor with 14 cores using Intel's TBB \cite{Reinders:2007} for parallelisation.  
We consider a regular base grid and allow a dynamic refinement
criterion to add up to two additional levels $\Delta \ell \in \{1,2\}$ of cells
around the shock front. In these tests all kernel level optimisations that are available in ExaHyPE have been used. As a result, scalability is more challenging, however, we are interested mainly in reducing the overall run-time.
 
The scalability improves as the work per element increases, i.e.~it improves
with higher polynomial degree $N$ or for increasing dimension. 
In general
good scalability can be observed on up to 14 cores at higher orders. 
All ExaHyPE codes employ a hybrid parallelisation strategy with
at least two MPI ranks per node \cite{CharrierUndEkaterinasPaper2019}. 
Results for this hybrid parallelisation strategy are provided in Section \ref{sec-strong-scaling}.

\subsection{Elastic Wave Equation}\label{sec-num-elast}
The restriction of ExaHyPE to Cartesian meshes seems restrictive in the context of most seismic applications. In this section, we introduce two methods for incorporating complex geometries into such a mesh {\color{black}\textit{without modifying the underlying numerical method}}. {\color{black} 
We would like to highlight that both of these methods omit manual mesh-generation, which is typically required as a pre-processing step in computational seismology applications and poses a major bottleneck. Hexahedral mesh generation can easily consume weeks to months and is limited for complex geometries of boundary and interface conditions, while form-fitted unstructured tetrahedral meshes allow for automatised meshing and complex geometries \cite{Wolherr,Ullrich}, however, pose numerical challenges, e.g. in form of misshaped sliver elements \cite{seissol}.}

\subsubsection{Diffuse Interface Approach}
\textcolor{black}{As in the Euler equations we can extend the elastic wave equation by a parameter $\alpha$, which reprents the volume fraction of the solid medium present in a control volume \cite{tavelli:dim}. }
Diffuse interfaces completely avoid the problem of mesh generation,
since all that is needed for the definition of the complex surface topography is to set a scalar colour  function to unity inside the regions covered by the solid and to zero outside.

The diffuse interface model is given by:
  \begin{displaymath}\begin{split}
  \frac{\partial\bm{\sigma}}{\partial t} - {E}(\lambda, \mu) \cdot \frac{1}{\alpha} \nabla {(\alpha\bm{v})} + \frac{1}{\alpha} {E}(\lambda, \mu) \cdot {\bm{v}} \otimes \nabla \alpha = 0, \nonumber \\
  \frac{\partial \alpha \bm{v}}{\partial t}-\frac{\alpha}{\rho}\nabla \cdot \bm{\sigma} - \frac{1}{\rho}\sigma\nabla \alpha=0, \nonumber \\ 
  \end{split}\end{displaymath}
  where ${E}(\lambda,\mu)$, $\lambda$, $\mu$, $\rho$ and the stress tensor $\bm{\sigma}$ are defined as in the  Section \ref{sec-elastic}.
  
  The material parameters are assumed to remain constant, i.e.
  \begin{displaymath}
     \frac{\partial\alpha}{\partial t} = 0, \qquad
     \frac{\partial\lambda}{\partial t} = 0, \qquad
     \frac{\partial\mu}{\partial t} = 0,
     \qquad \frac{\partial\rho}{\partial t} = 0.
  \end{displaymath}  
Physically, $\alpha$ represents the volume fraction of the solid medium present 
in a control volume. The equations become non-linear in those regions in which $0<\alpha<1$.

This formulation can be extended to allowing moving materials. Instead of solving $\frac{d\alpha}{dt}=0$, we can solve $$\frac{d\alpha}{dt}+v\nabla \alpha=0.$$ In this way the free surface boundary is allowed to move according the local velocity field \cite{tavelli:dim}.

To verify the accuracy of this method we solve the layer over homogeneous halfspace (LOH.1) benchmark problem described by Day et al. \cite{Day:LOH1}. 
This problem is a well-known reference benchmark for seismic wave propagation in numerical codes. 
The LOH.1 benchmark considers way propagation in a hexahedral geometry filled
  with two materials that are stacked on top of each other. The first material
  is characterised by a lower density and smaller seismic wave speeds. The exact material parameters are defined in Table \ref{tbl:loh1}. The parameters $\lambda$ and $\mu$ of the equation can be derived from these values.

\begin{figure}[t]
        \centering{
	\includegraphics[width=0.4\textwidth]{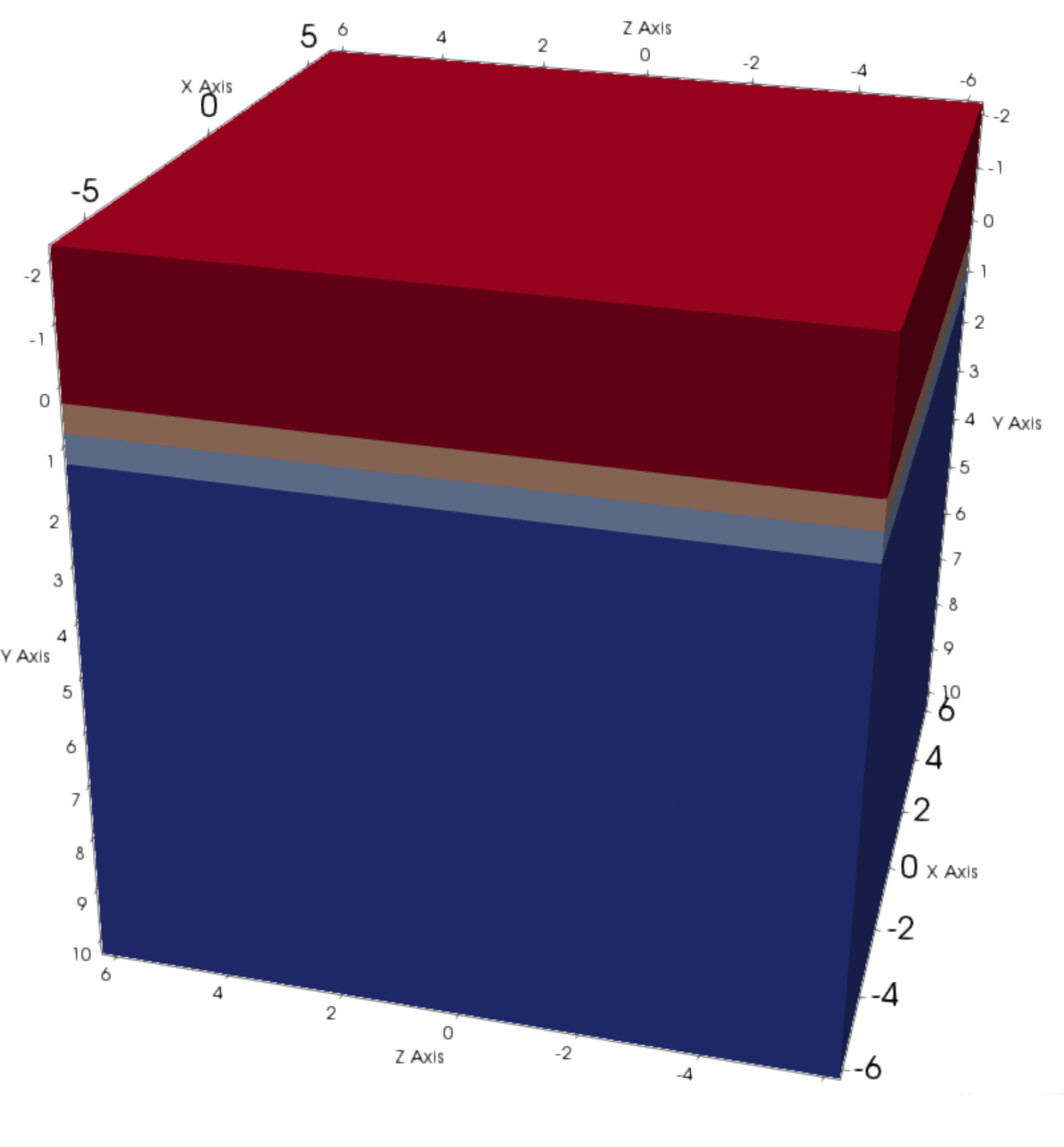}
	\hspace{0.05\textwidth}
	\includegraphics[width=0.4\textwidth]{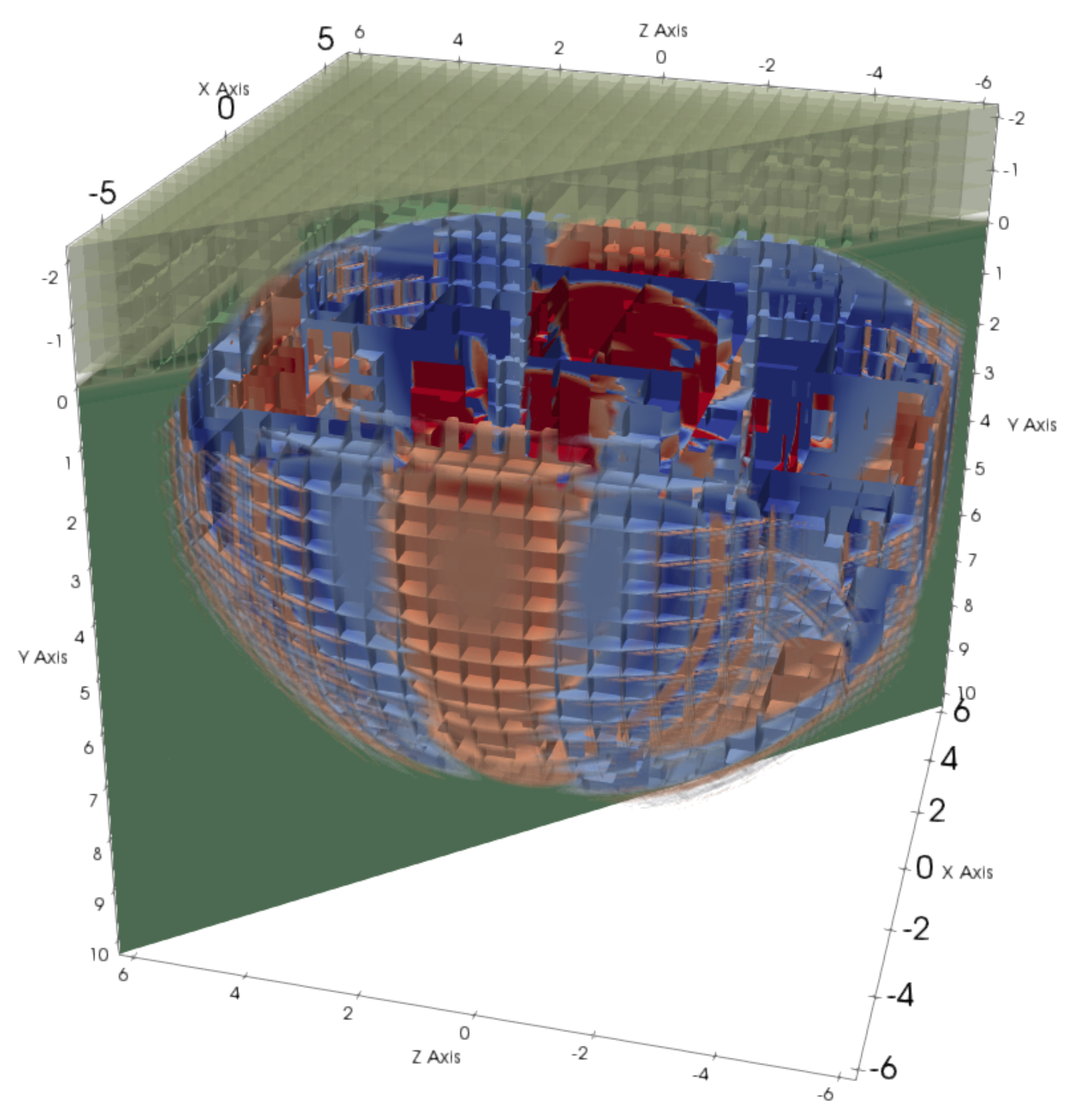}
        }
	\caption{From left to right: Plot of the limited area indicated by $\alpha$; The velocity field in $x$ with the simulated topography in the background at $t=1.1$(transparent indicates the free surface).
	}
	\label{fig-dim}
\end{figure}

  \begin{table}[bh]
    \centering
    \caption{Material parameters of the LOH.1 benchmark.}
    \label{tbl:loh1}
    \begin{tabular}{cccc}
       \toprule
      $x$ & $c_p[m/s]$ & $c_s[m/s]$ & $\rho[kg/m^3]$ \\
       \midrule
      $ < 1km$ & $4000$ & $2000$ & $2600$ \\ 
      $ \ge 1km$ & $6000$ & $3464$ & $2700$ \\ 
       \bottomrule
    \end{tabular}
  \end{table}

  A point source is placed $2$ km below the surface at the centre of the domain, such that the resulting wave propagates through the change of material.
  
  In the diffuse interface method we limit the free surface to be able to resolve the local discontinuity of $\alpha$. The left picture in Figure \ref{fig-dim} shows the resulting distribution of limited (red) and unlimited areas (blue) and indicates the transition of both (light blue and red). The right picture shows the velocity field in x direction at $t=1.1$. 
  Our implementation of the diffuse interface method is able to successfully resolve both the changing material parameters and the absorbing surface boundary conditions.

\begin{figure}[bt]
\centering
  \includegraphics[width=0.4\textwidth]{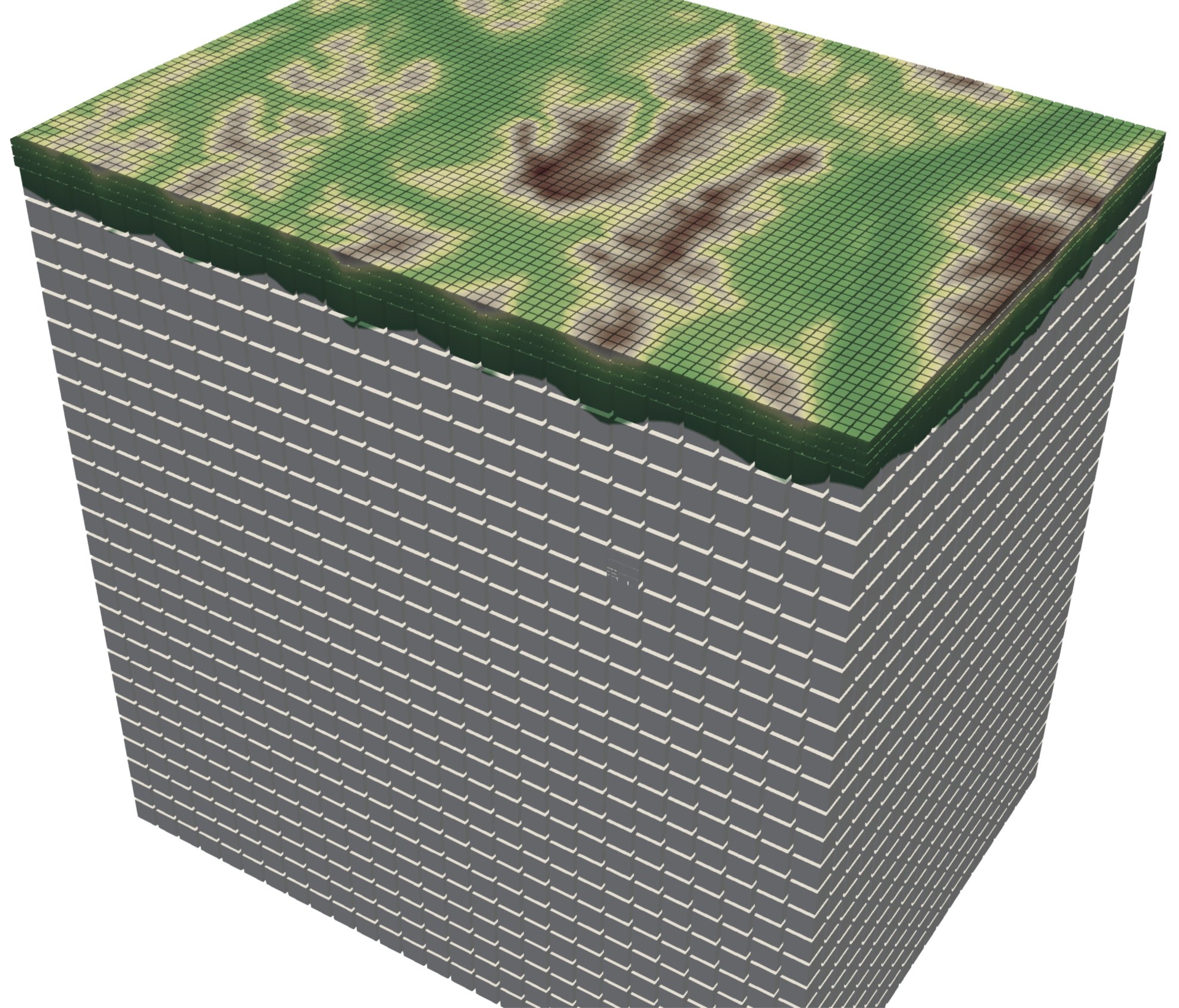}
  \hspace{0.05\textwidth}
  \includegraphics[width=0.4\textwidth]{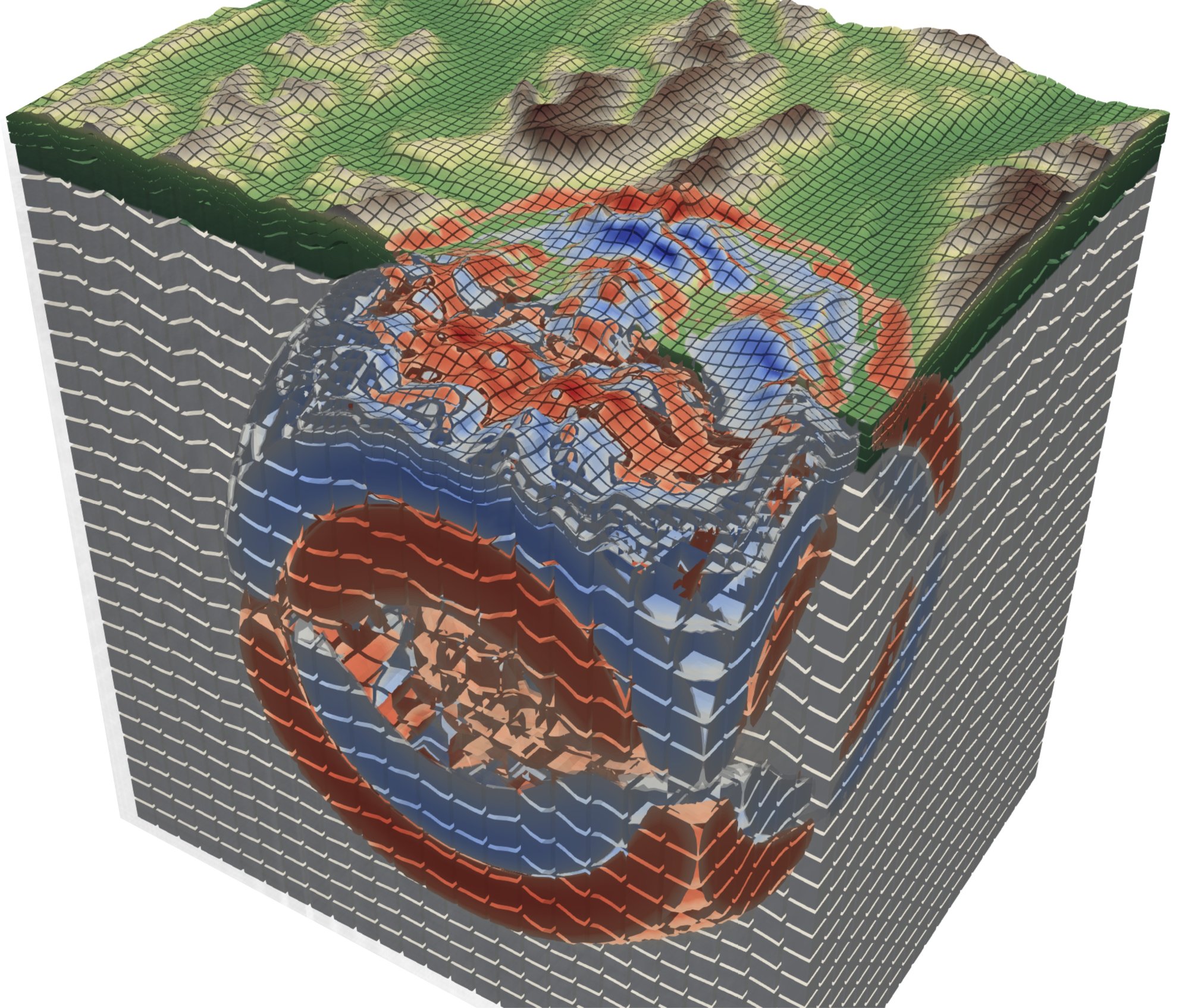}
  \caption{
  Simulation of seismic waves around the Mount Zugspitze with curvilinear meshes. The internal Cartesian mesh (left) is used to simulate a complex topography (right).}
  \label{fig-curvilinear}
\end{figure}

\subsubsection{Curvilinear meshes}
Our second approach models complex topographies by applying a curvilinear transformation to the elements of an adaptive Cartesian mesh. {\color{black} To do so we generate surface quadrature nodes depending on the topography, create a  2D curvilinear interpolation of those quadrature nodes on domain boundaries with topography curves and domain edges as constraints, and finally generate a Jacobian mapping $J$ at each node.} 
  Taking this mapping into account, the linear elastic wave equations
  can be written as:
  \begin{eqnarray}
    \frac{\partial \sigma }{\partial x} = \frac{1}{ J}\left(\frac{\partial }{\partial q}\left(Jq_x \sigma\right) + \frac{\partial }{\partial r}\left(Jr_x \sigma\right) + \frac{\partial }{\partial s}\left(Js_x \sigma\right)\right) \nonumber \\
    \frac{\partial v}{\partial x} = q_x\frac{\partial v}{\partial q} + r_x\frac{\partial v}{\partial r}  + s_x\frac{\partial v}{\partial s} \nonumber
  \end{eqnarray}
  where $J$ is the Jacobian matrix of the mapping from mesh to topography. This approach allows us to model meshes with complex topographies including faults and inner branches \cite{duru:curvilinear1, duru:curvilinear2}.
  {\color{black} The ADER-DG algorithm as described previously, can then be applied directly to this  version of the elastic wave equation.}
  
  Figure \ref{fig-curvilinear} shows a snapshot of a numerical experiment that simulates propagation of seismic waves in the area
around the mountain Zugspitze, in Germany. 
  Both the curvilinear approach and the diffuse interface method are able to resolve the complicated topographies in this scenario which is motivated by the AlpArray experiment.

\begin{figure}[tb]
\centering
\includegraphics[width=0.45\textwidth]{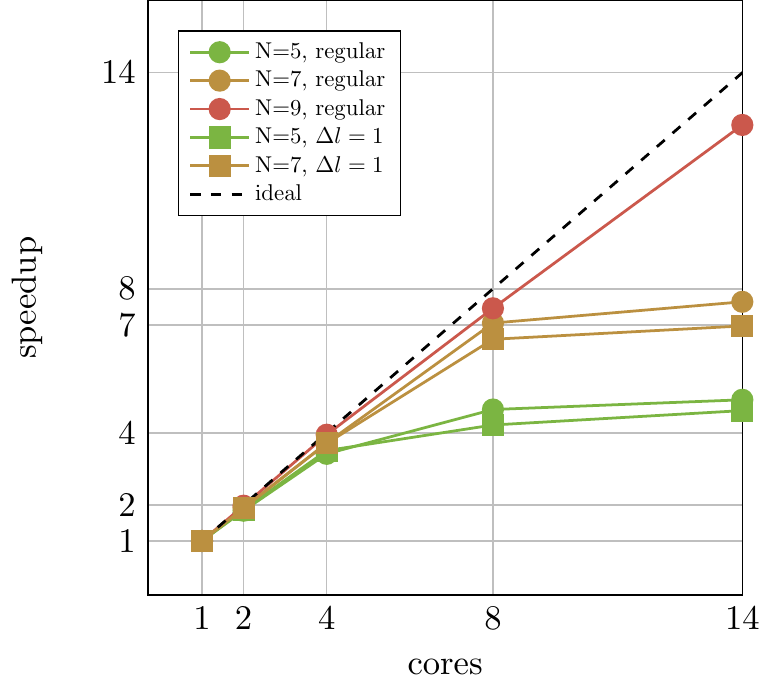}
\hspace{0.05\textwidth}
\includegraphics[width=0.45\textwidth]{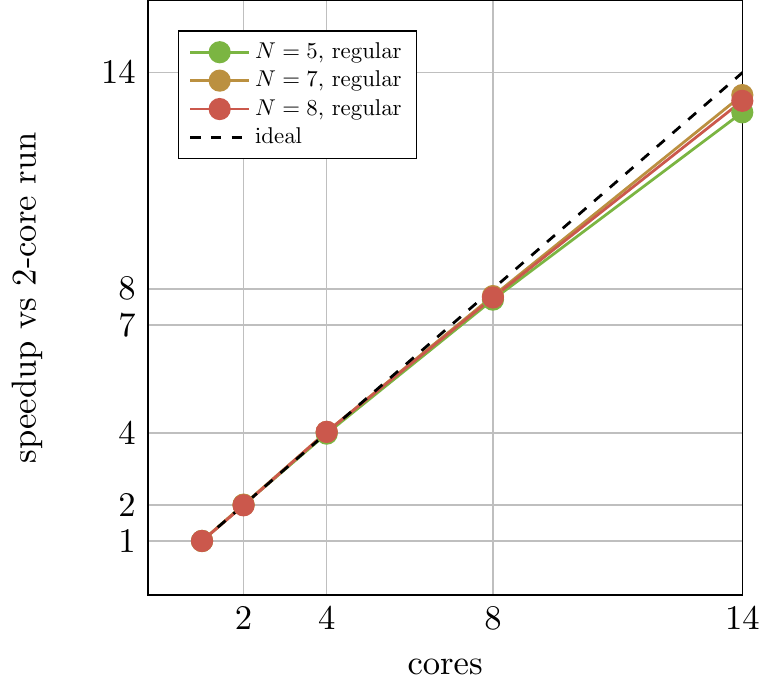}
\caption{Shared-memory scalability of the linear ADER-DG implementation on SuperMUC's Phase 2 when running the LOH.1 benchmark.
Left: Curvilinear elements are used. Right: The diffuse interface method is used.
}
\label{fig-elastic-loh1-tbb}
\end{figure}

Figure \ref{fig-elastic-loh1-tbb} shows the shared-memory scalability of the linear ADER-DG implementation on SuperMUC's Phase 2 when running the LOH.1 benchmark comparing the scalability of the two meshing approaches.
We consider a regular base grid with $27^3$  cells and allow a dynamic refinement criterion to add one additional level of cells to the layer over the halfspace. Memory constraints prevented further refinement. As in the non-linear test case the scalability improves as the work per element increases. However, due to the lower amount of total work in the linear setting a higher polynomial degree needs to be reached to attain scalability in this case. This means that the overall scalability of the diffuse interface approach is higher, however, this comes at the price of a higher overall computational cost due to the non-linearity of the formulation. 

The question of which method should be used is highly problem dependent. 
The curvilinear method is purely linear and as such computationally cheap. The non-linear diffuse interface approach, on the other hand, requires additional limiting on the surface. However, the time step size for the DIM is independent of the simulated topography, while in the curvilinear method it is highly dependent on the distortion introduced by the topography. Simulations with a relatively smooth, flat surface are expected to be faster with the curvilinear method, while highly varying topographies are better discretised with the DIM. DIM has the added advantage of allowing moving free surface boundaries, a feature that is difficult to realise with curvilinear meshes.

\subsection{Shallow Water Equations}

We demonstrate the dynamic mesh refinement capabilities of ExaHyPE
  via the shallow water equations \eqref{eq-swe}.
Even more importantly, we present a non-toy problem which uses ExaHyPE's 2D
facilities.
To solve  \eqref{eq-swe}, we use the a-posteriori limiting ADER-DG method and
  equip its FV limiter with a recently developed HLLEM Riemann
  solver \cite{Dumbser:2016:HLLEM}. The solver allows for wetting and drying.
This also demonstrates that users can inject their own Riemann solvers in ExaHyPE.

The same scenario is available in ExaHyPE using a Finite Volume scheme instead of an ADER-DG  scheme with FV limiter \cite{rannabauer:swe:2018,Rannabauer:2018}.
In both implementations the same Riemann solver can be used. The Riemann solver considers a dry tolerance $\epsilon>0$ below which cells are marked as \textit{dry}.
The constant $\epsilon$ is necessary to avoid negative water heights, which usually lead to unphysical non-linear effects. 
If one of the cells is flooded the jump in bathymetry is taken into account by the Riemann solver. 
When reconstructing the DG solution
from the FV limiter the water level is reconstructed first and then the bathymetry is subtracted. This ensures that the non-linear reconstruction process does not produce artificial waves.  
In tsunami simulations, this scheme allows us to simulate areas close to the coast with the FV subcell limiter, while areas in the deep ocean are processed by the high order ADER-DG method.

\begin{figure}[tb]
\centering
  $t = 0$  \hspace{0.23\textwidth}  $t=1$  \hspace{0.23\textwidth}  $t= 2$\\   
 \includegraphics[width=0.32\textwidth]{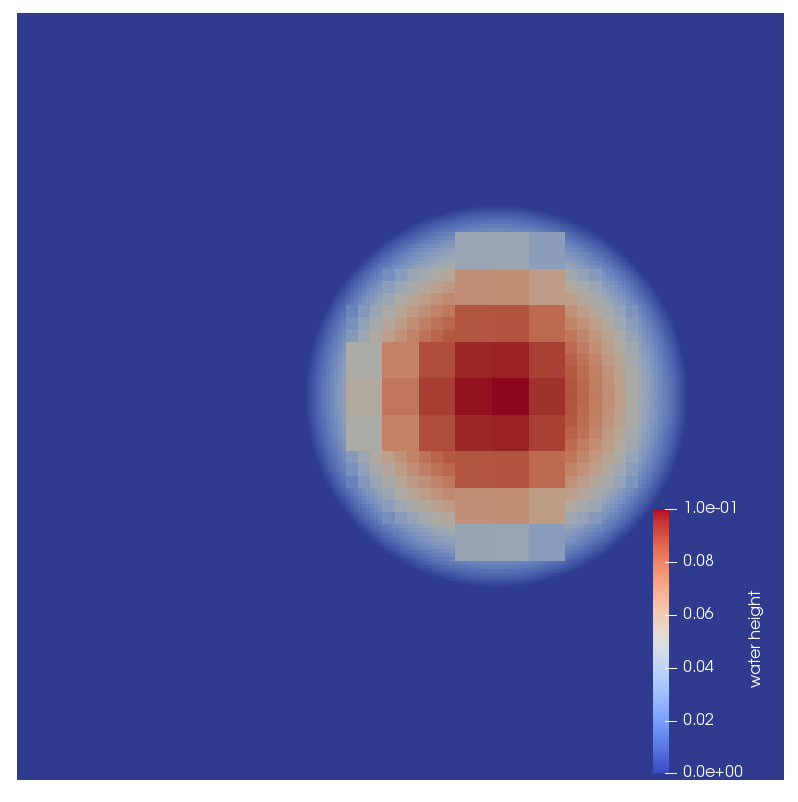}
 \includegraphics[width=0.32\textwidth]{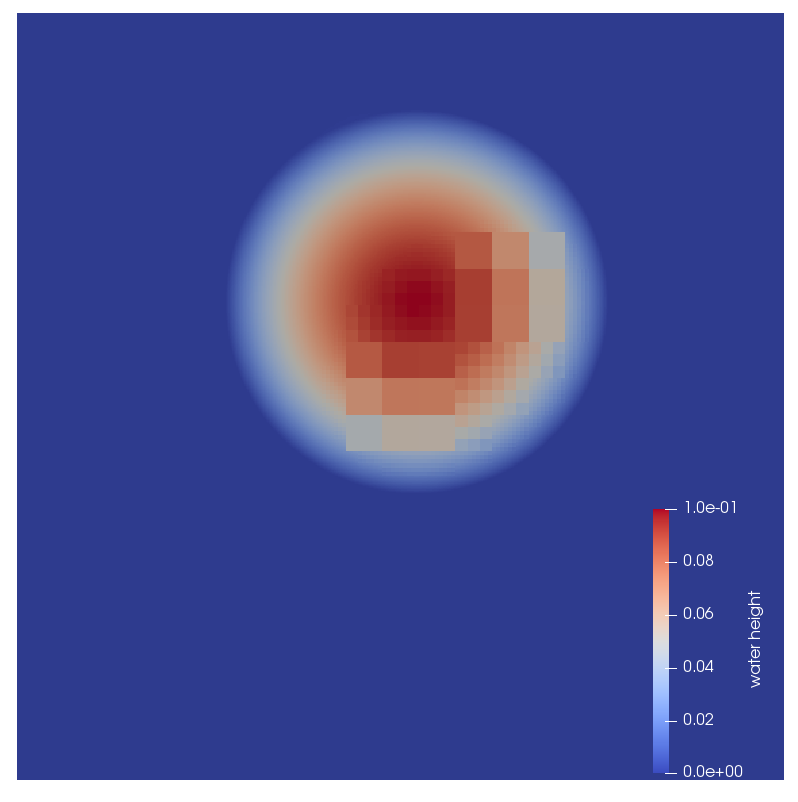}
 \includegraphics[width=0.32\textwidth]{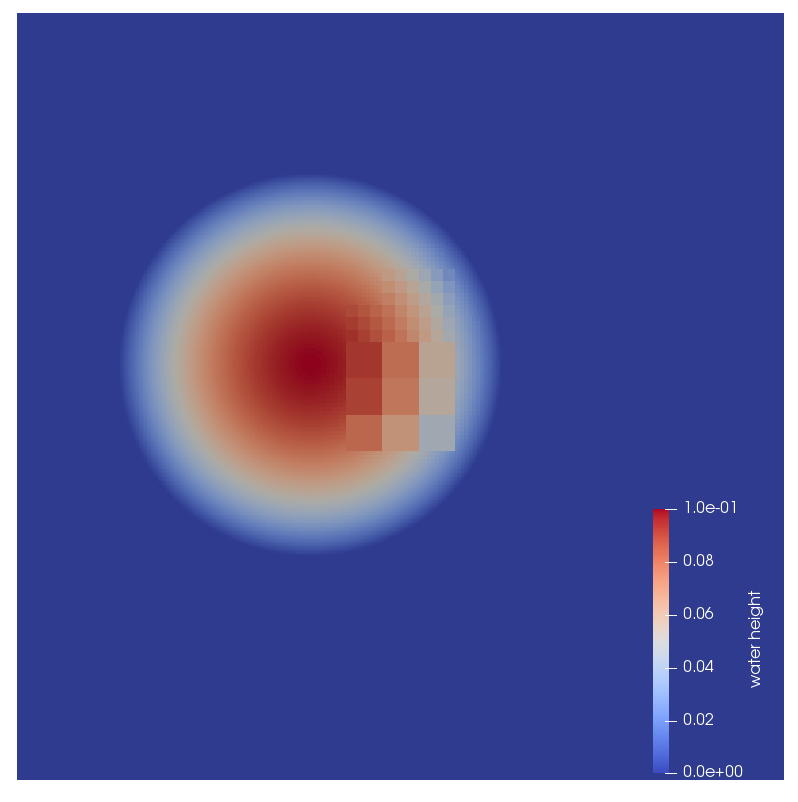}\\
 \includegraphics[width=0.32\textwidth]{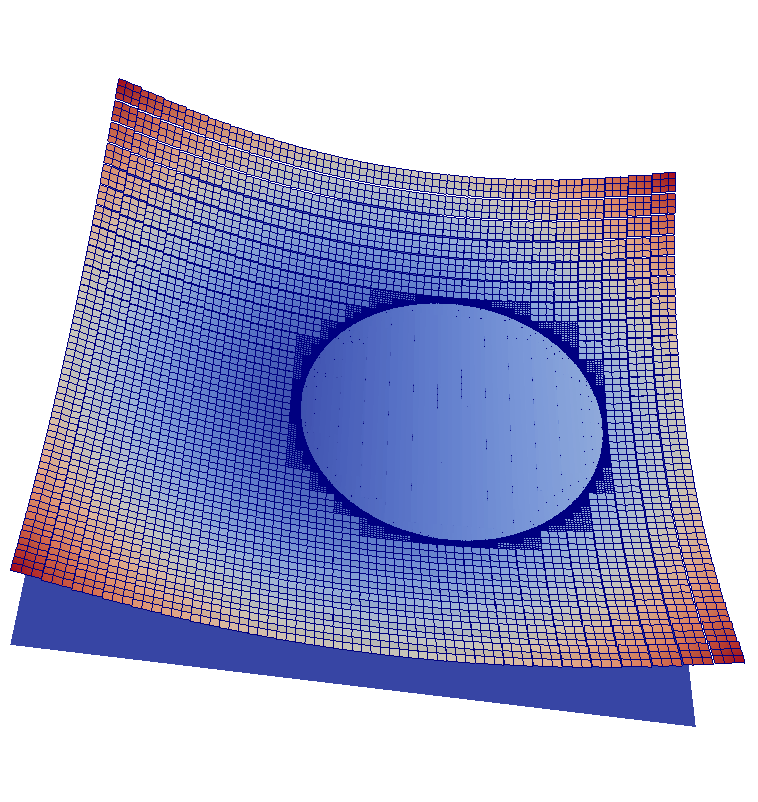}
 \includegraphics[width=0.32\textwidth]{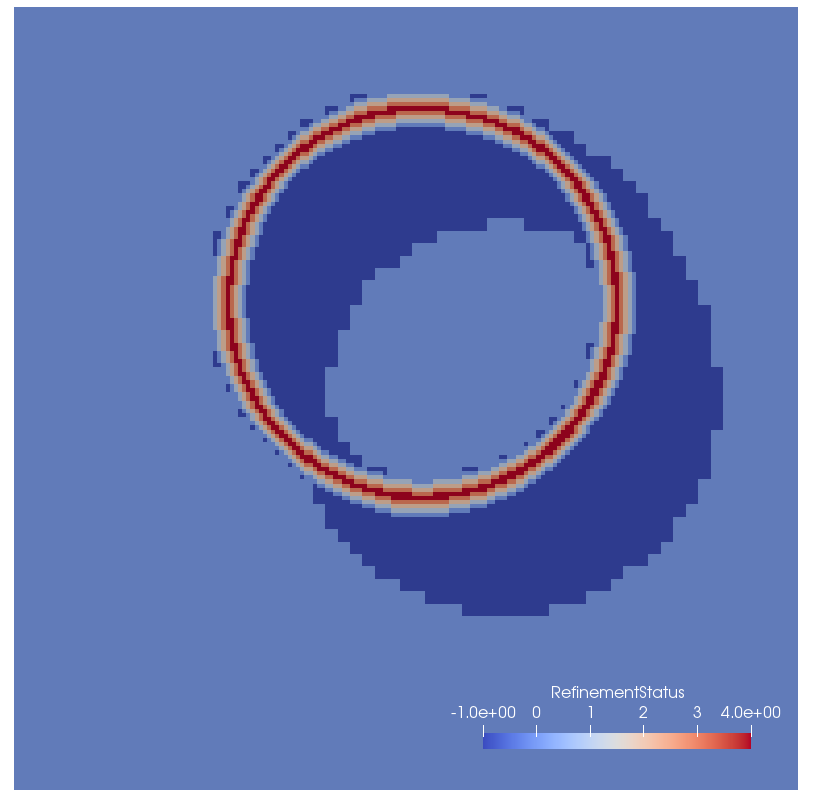}
 \includegraphics[width=0.32\textwidth]{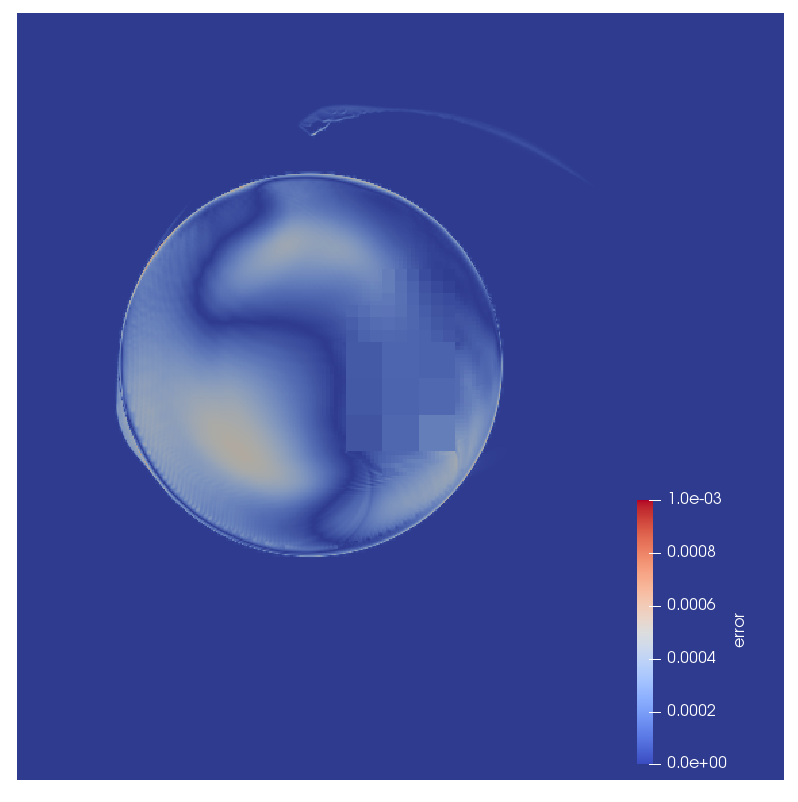}

 \caption{ADER-DG with a-posteriori limiting for the oscillating lake scenario (shallow water equations). Top row: water height at time $t=0,1,2,3$. Bottom row: The locations (red) in which the FV limiter is active, the adaptively refined mesh and the error at times $t=2$ and $3$.}
 \label{fig-swe}
\end{figure}

To test the shallow water equations given in \eqref{eq-swe} we use the oscillating lake scenario. Here, a water droplet travels in circular motion over a dry basin. {\color{black} The topography of the basin is resolved using the final variable $b$, which remains constant throughout the simulation.}  The analytical solution, which is used as an initial condition is given by
\begin{displaymath}
 Q = \begin{pmatrix}
      h\\ hu \\ hv\\ b
     \end{pmatrix}
  =
  \begin{pmatrix}
     \max\left\{0, \frac{1}{10} \left(x_0  \cos(\omega \cdot t) + x_1 \sin(\omega \cdot t) + \frac{3}{4}\right) - b\right\}\\
   \frac{1}{2} \omega \sin(\omega \cdot t) h\\[.1em]
    \frac{1}{2}\omega \cos(\omega \cdot t) h\\
    \frac{1}{10} (x_0^2 + x_1^2)
    \end{pmatrix},
\end{displaymath}
where $ \omega = \sqrt{0.2 g}$ and $g=9.81$ denotes the gravitational constant. The initial conditions are supplemented by wall boundary conditions. In this setup we use polynomial order $N=3$ and a base grid of $7\times7$ elements. We adaptively refine mesh elements where the water height is small but not zero.
  We allow up to two levels of adaptive refinement. 

This scenario provides a challenging test case for numerical codes due to the continuous wetting and drying of cells. There exists an analytical solution for this benchmark scenario allowing us to verify the well-balancedness of a scheme, as well as in the resolution of drying or inundated cells. 

In the top row of Figure \ref{fig-swe}, we plot the water height at $t=0,1,2,3$. Despite the frequent wetting and drying the scheme performs well. In those areas in which the DG solution is used, we note the accuracy of the higher order scheme on a coarse grid. In those areas in which a FV solution is necessary, the dynamic AMR is useful to retain accuracy and avoid diffusion.
The bottom row of Figure \ref{fig-swe} shows the locations in which the FV limiter is active and adaptive mesh refinement is used. Furthermore, it shows the error in the water height. We observe that the FV limiter is used only in those cells in which it is required, the limiter locations move with the solution. The error remains below $10^{-3}$ in all cells. 

\begin{figure}[tb]
\centering
\includegraphics[width=0.45\textwidth]{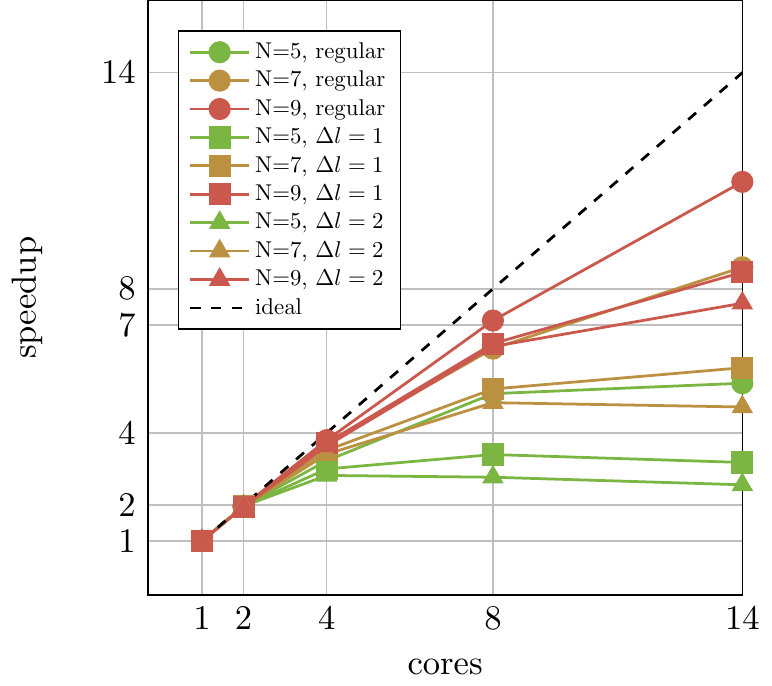}
\caption{ Shared-memory scalability of the non-linear ADER-DG implementation for the oscillating lake experiments using a base grid of $243^2$ cells with up to two levels of adaptive refinement.}
\label{fig-swe-tbb}
\end{figure}

 In Figure \ref{fig-swe-tbb}, we show the shared-memory scalability for the oscillating lake. This scaling test was run on one $2.2$ GHz $14$ core Intel Xeon E5-v3 processor of SuperMUC Phase~2.  We consider a regular base grid and allow a dynamic refinement criterion to add up to two additional levels of cells around the water droplet. In this area the FV limiter is active.  As in the previous test case the scalability improves for higher polynomial degree $N$. 

\subsection{General Relativistic Magneto-Hydrodynamics}\label{sec-num-grmhd}

\begin{figure}[tb]
\centering
 \includegraphics[width=0.45\textwidth]{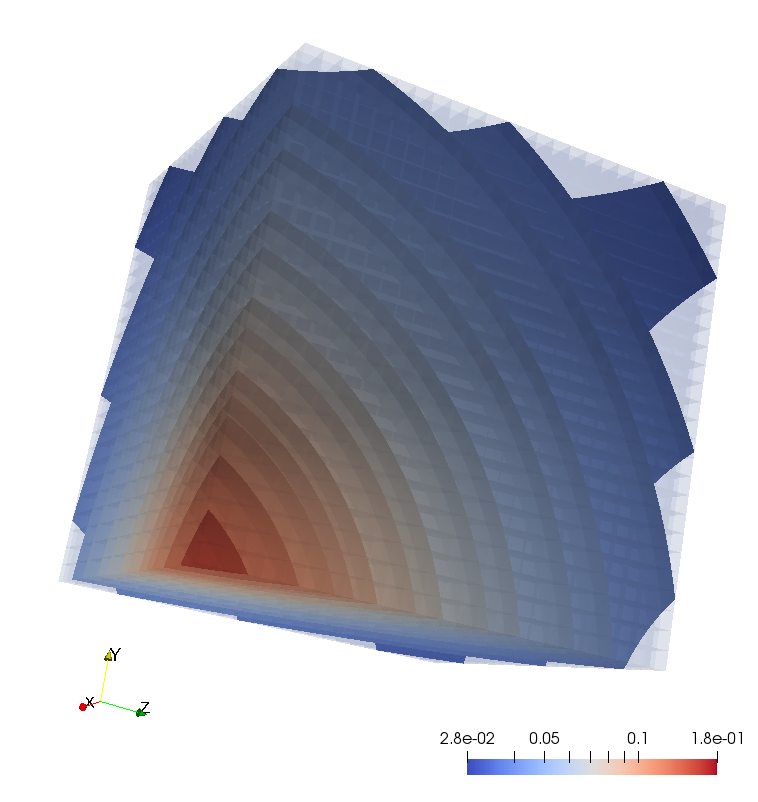}
 \includegraphics[width=0.45\textwidth]{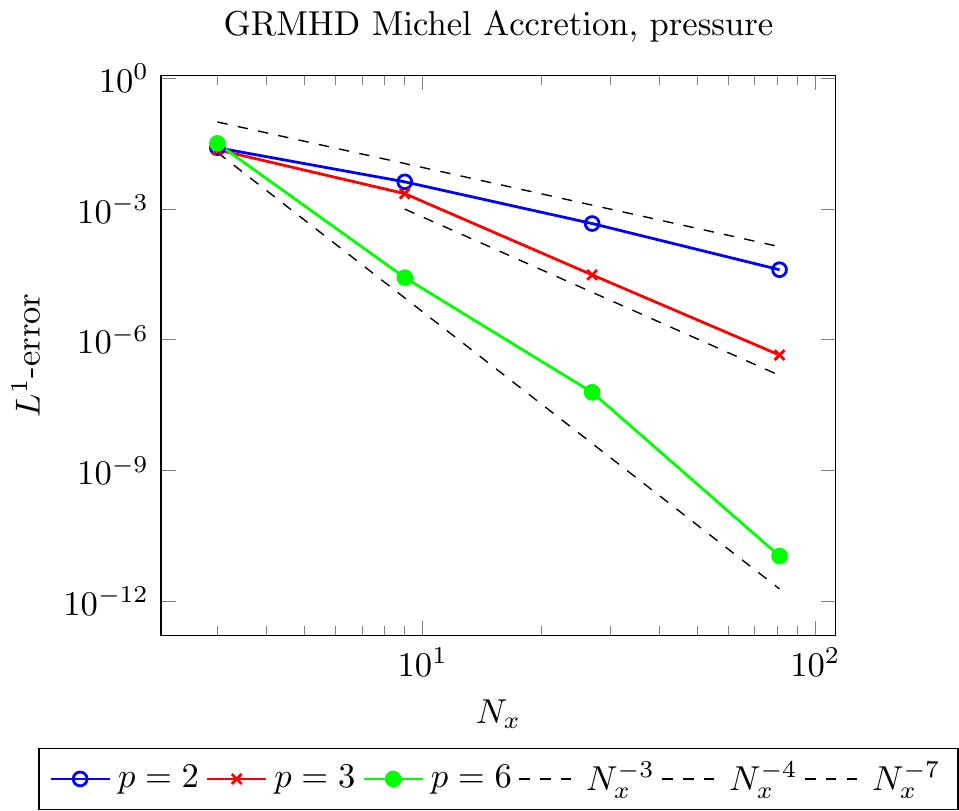}
 \caption{(Left) The solution  for pressure of the Michel accretion test at $t=1.0$. (Right) Convergence of the ADER-DG scheme at $t=1.0$. The dotted lines show the expected convergence rates.}
\label{fig-michel}
 \end{figure}

\begin{table}[b]
\centering
\caption{Convergence of the ADER-DG scheme for the Michel accretion test at $t=1.0$. }
\label{table-michel}
\begin{tabular}{ccccccc}
 \toprule
 & \multicolumn{2}{c}{$N = 2$} & \multicolumn{2}{c}{$N = 3$} & \multicolumn{2}{c}{$N = 6$} \\
 \cmidrule(lr){2-3}
 \cmidrule(lr){4-5}
 \cmidrule(lr){6-7}
 \#cells & Error $L^1$ & Order & Error $L^1$ & Order & Error $L^1$ & Order\\
 \midrule
$3^3$  &	2.56e-02 &       &	2.31e-02 &          &	3.23e-02 &	\\
$9^3$  &	4.22e-03 & 	1.65 &	2.27e-03 &	2.11    &	2.28e-05 &	 6.45\\
$27^3$ & 	4.66e-04 &	2.01 &	3.09e-05 &	3.91    &	6.24e-08 &	 5.52\\
$81^3$ &	4.06e-05 &	2.21 &	4.47e-07 &	3.86    &	1.09e-11 &	 7.88\\
 \bottomrule
\end{tabular}
\end{table}

To uncover and validate ExaHyPE's high convergence order, we
require a test case which does not contain discontinuities, as the limiter
reduces the convergence order.
Furthermore, the analytical solution has to be known. 
We simulate the spherical accretion
onto a stationary black hole. 
This well-known test case has a known analytical solution \cite{michel:1972}. 
Details of the setup in the context of ADER-DG methods can also be found in \cite{fambri:2017}. This setup is challenging since 
GRMHD is a non-linear equation of $19$ variables containing both fluxes and non-trivial non-conservative products. 
Further complexity  arises from the the closeness of the singularity at the critical radius to the computational domain.

We use the analytical solution as the external state vector for the Riemann solver at the boundary of the domain $\partial \Omega$. The test is performed in Kerr-Schild coordinates on the spatial domain $[2.2,12.2]^3$, away from the critical radius.

In Figure \ref{fig-michel} we show the solution at time $t=1.0$ and the convergence of the error in terms of $L^1$-norm for three different polynomial degrees $N=2,3$ and $6$. 
Table \ref{table-michel} lists $L^1$-errors and convergence order for each polynomial degree. This  shows that our scheme converges to the expected order of $N+1$.  

 For the next test, we move to a simulation for which the analytical solution is not known.
In this test, a stable non-rotating and non-magnetised neutron star is simulated in three space dimensions by solving the GRMHD equations in the Cowling approximation, i.e. assuming a static background spacetime. The initial state has been obtained as a solution to the Tolman-Oppenheimer-Volkoff (TOV) equations. The corresponding fluid and metric variables are compatible with the Einstein field equations. We set the magnetic field to zero for TOV stars.

The TOV equations constitute a non-linear ODE system in the radial space coordinate, that has been solved numerically by means of a fourth order Runge-Kutta scheme on a very fine grid with step size $dr=0.001$. The parameters of the problem have been chosen to be: $\rho_c=1.28 \cdot 10^{-3}$, adiabatic exponent $\Gamma=2$ and a constant atmospheric pressure $p_{\text{atm}}=10^{-12}$. 

 The star sits at the origin of the domain {\color{black} $ [-15,15]^3$}. We apply reflection boundary conditions at the three simulation boundary surfaces $x=0$, $y=0$ and $z=0$. At the surfaces $x=R$, $y=R$ and $z=R$, we apply exact boundary conditions, evaluating the initial data at the boundary.

\begin{figure}[h]
\centering
\includegraphics[height=5cm]{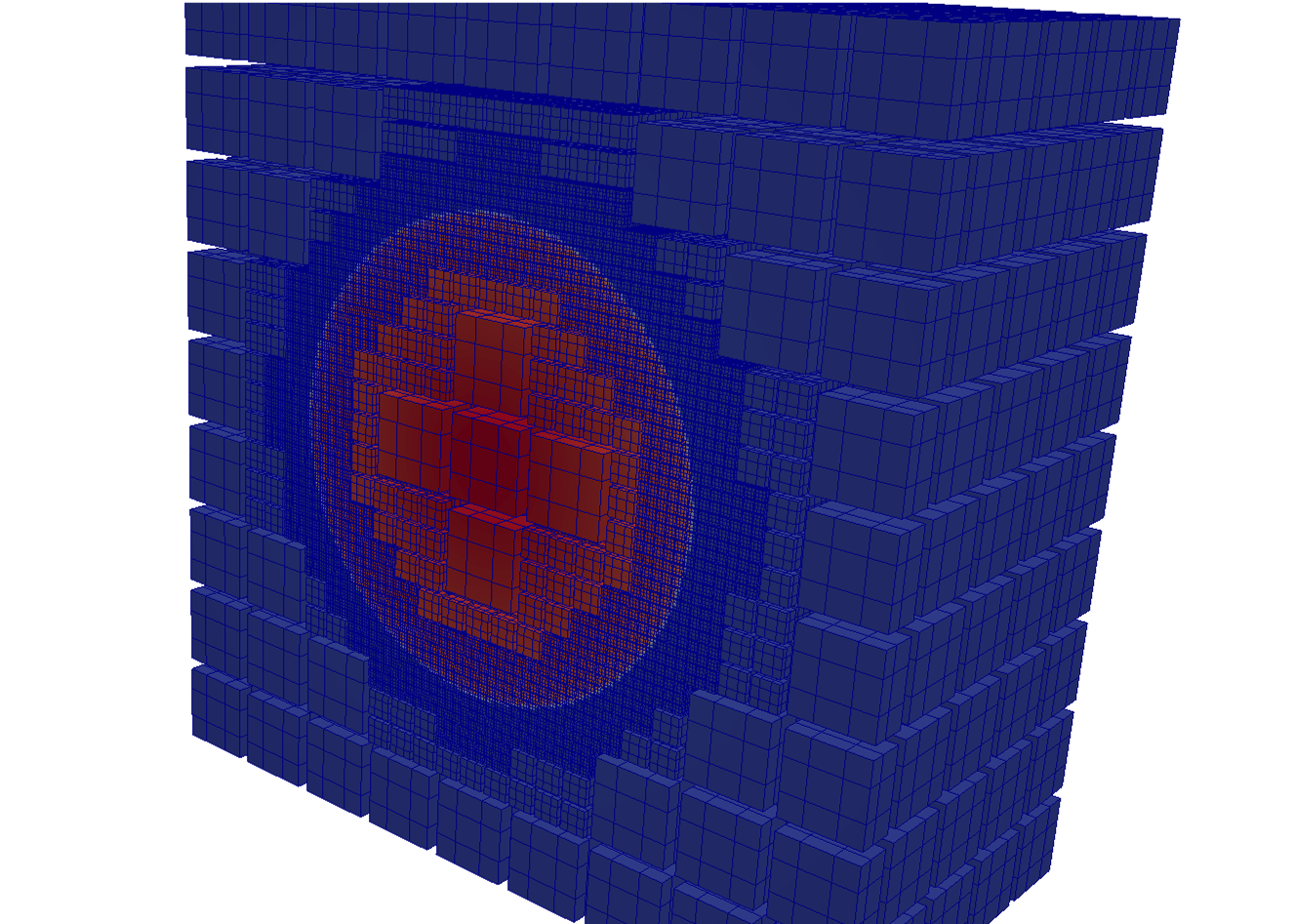}
\includegraphics[height=5cm]{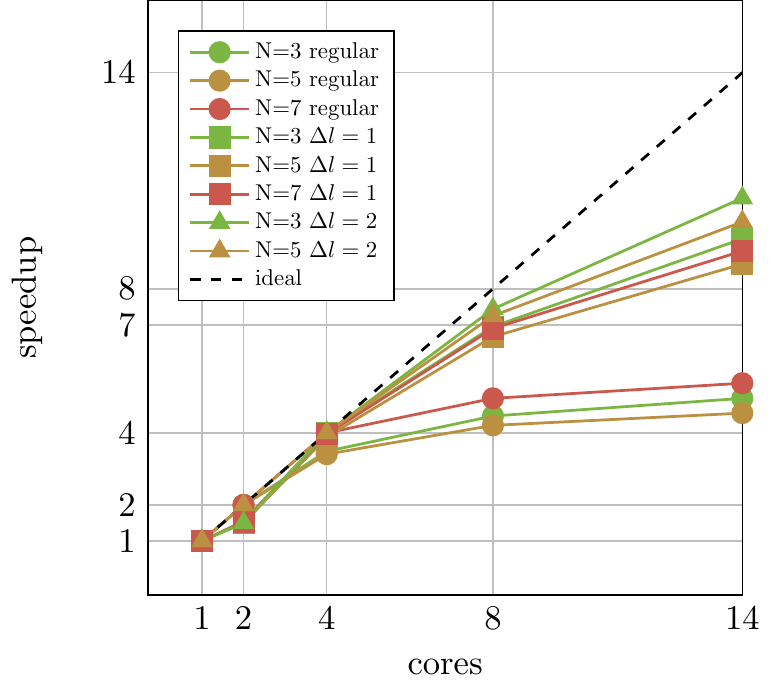}
\caption{Left: The TOV star setup showing two levels of adaptive mesh refinement along a sphere around the star. Right: Shared-memory scalability of the non-linear hybrid ADER-DG - FV implementation  when running the TOV star example. }
\label{fig-grmhd-tbb}
\end{figure}

In Figure \ref{fig-grmhd-tbb}, we show the shared memory of scalability of the the non-linear hybrid ADER-DG - FV implementation of the GRMHD equations for the TOV star run on SuperMUC phase 2. We consider a regular base grid with $27^3$ cells and allow a dynamic refinement criterion to add one additional level of cells along a sphere around the TOV star.

\subsection{Compressible Navier-Stokes}
The compressible Navier-Stokes equations~(\ref{eq:navier-stokes}) demonstrate that ExaHyPE is not only able to solve hyperbolic PDEs, but is also capable of handling viscous effects. We use the numerical flux of~\cite{Gassner:2008:ViscousDG} to integrate this into our ADER-DG framework.

\begin{figure}[bt]
\centering
\includegraphics[width=0.5\textwidth]{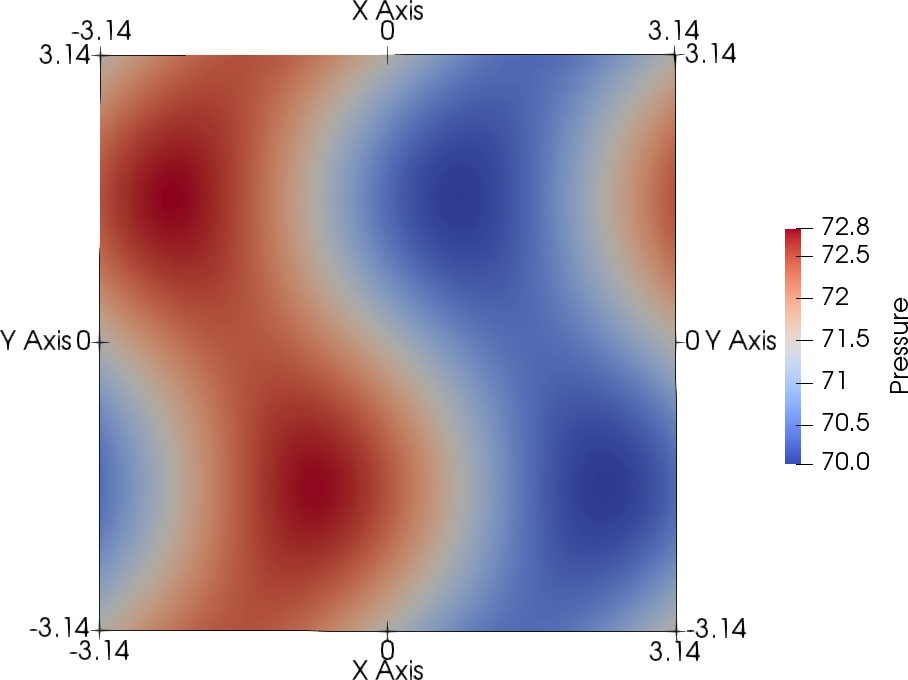}~
\includegraphics[width=0.5\textwidth]{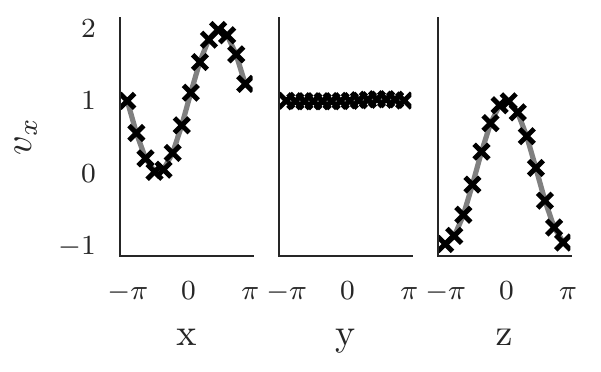}
\caption{
Left: Pressure of the ABC-flow, $x$-$y$ slice.
  Right: Velocity slices of the ABC-flow. Markers indicate samples of our solution, line indicates analytical solution. Image reproduced from~\cite{Krenz:19:ExaCloud}
}
\label{fig:ns-abc}
\end{figure}

To test the equations, we utilise the Arnold-Beltrami-Childress (ABC) flow
\begin{align*}
  \label{eq:abc-flow}
  \begin{split}
  \rho (x, t) &= 1,\\
  v (x, t) &= \phantom{-} \exp(-1\mu t)
  \begin{pmatrix}
    \sin(z) + \cos(y)\\
    \sin(x) + \cos(z)\\
    \sin(y) + \cos(x)
  \end{pmatrix}, \\
  p(x, t) &= -\exp(-2 \mu t) \, \left(\cos(x)\sin(y) + \sin(x)\cos(z) + \sin(z)\cos(y)\right)
  + C,
  \end{split}
\end{align*}
 where the constant $C = 100/1.4$ governs the Mach number and $\mu = 0.01$ is the viscosity~\cite{Tavelli:2016:StaggeredDG}. This equation has an analytical solution in the incompressible limit. We impose this solution at the boundary.
 In Figure~\ref{fig:ns-abc} we show a comparison between the analytical solution and a simulation with order $N=2$ on a regular mesh of $27^3$ cells. We see a good agreement with the analytical solution a time $t = 1.0$.

\begin{figure}[tb]
\centering
\includegraphics[width=0.55\textwidth]{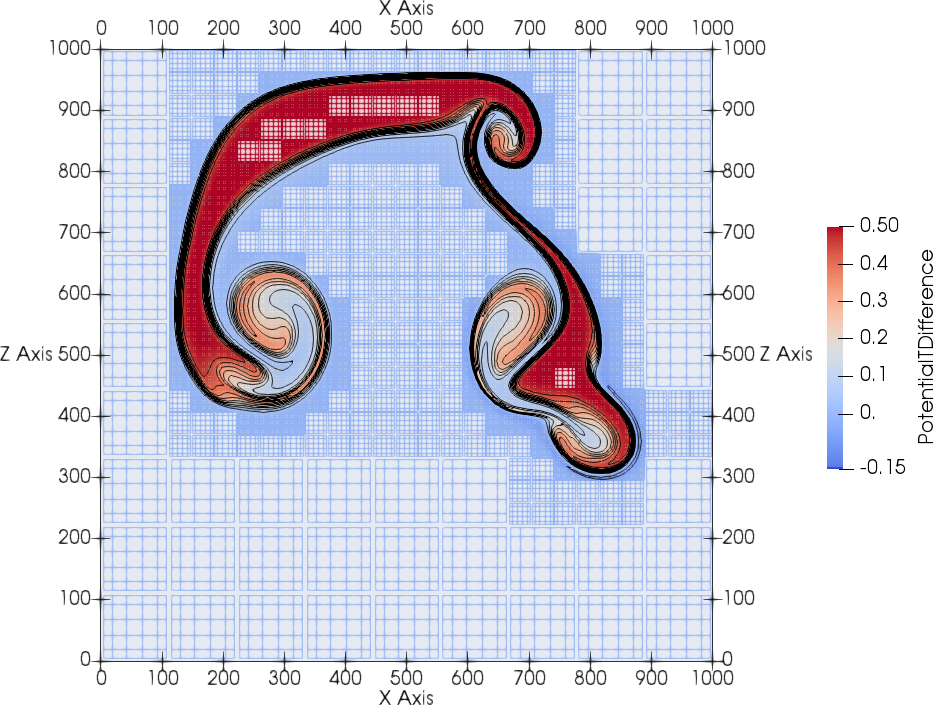}~
\includegraphics[width=0.43\textwidth]{{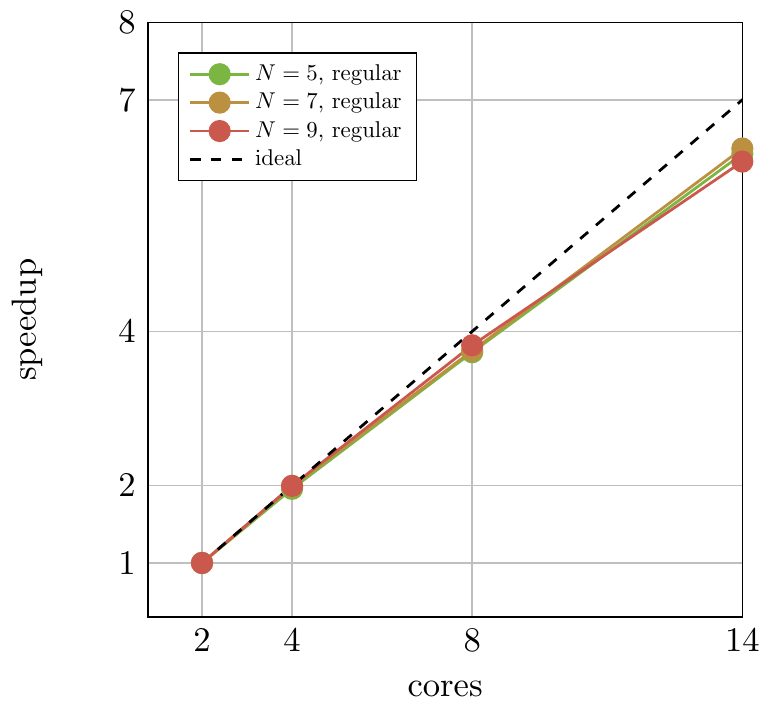}}
\caption{Left: The colliding bubble scenario with two levels of dynamic adaptive mesh refinement. Image taken from~\cite{Krenz:19:ExaCloud}
Right: Shared-memory scalability of the nonlinear ADER-DG implementation when running the ABC-flow example with viscous effects.}
\label{fig:ns-abc2}
\end{figure}

In addition, we evaluate our scenario for the colliding bubbles scenario of~\cite{Muller:2010:Cumulus}. The initial conditions of this scenario are obtained by an atmosphere with a background state that is in hydrostatic balance, i.e. where
\begin{equation*}
  \label{eq:hydrostatic-balance}
  \frac{\partial}{\partial z} p{\left (z \right )} = -g \rho (z).
\end{equation*}
Initially, the domain has the same potential temperature $\Theta$.
This is then perturbed by 
\begin{equation*}
  \label{eq:bubbles-pertubation}
  \Theta' =
  \begin{cases}
    A & r \leq a, \\
    A \exp \left( - \frac{(r-a)^2}{s^2} \right) & r > a,
    \end{cases}
\end{equation*}
where $r$  is the Euclidean distance from the centre of the bubble $(x_c, z_c)$ and the spatial position $(x, z)$.
We have two bubbles, with constants
\begin{align*}
\begin{alignedat}{6}
  & \text{warm:} \qquad && A = \SI{0.5}{}, \quad&& a = \SI{150}{\m}, \quad&& s = \SI{50}{\m}, \quad&& x_c = \SI{500}{\m,} \quad&& z_c = \SI{300}{\m},\\
  & \text{cold:} \qquad && A = \SI{-0.15}{}, \quad&& a = \SI{0}{\m}, \quad&& s = \SI{50}{\m}, \quad&& x_c = \SI{560}{\m}, \quad&& z_c = \SI{640}{\m}.
  \end{alignedat} 
  \end{align*}
The simulation runs for \SI{600}{\s} and uses a viscosity of $\mu = 0.01$ for regularisation. We use order $N=5$ and a mesh with up to two adaptive AMR levels.
The results of this simulation are shown in Figure~\ref{fig:ns-abc2} (left). There is an excellent agreement with the reference solution of~\cite{Muller:2010:Cumulus}.
For further details on the setup, we refer to~\cite{Krenz:19:ExaCloud}.
In Figure~\ref{fig:ns-abc2} (right), we show the shared-memory scalability using the ABC-flow on SuperMUC phase 2. In this test case the shared memory scalability is very good for all tested polynomial orders. 

\subsection{Hybrid Parallelisation Strategy}\label{sec-strong-scaling}
In the previous sections we showed the shared memory scalability of various test cases in
two and three dimensions. In general we observed that the scalability improves as the work per element increases, i.e.~it improves with higher polynomial degree $N$, for increasing dimension, or PDE complexity. In this section we demonstrate the effectiveness of a hybrid parallelisation strategy using the GRMHD TOV star setup described in Section \ref{sec-num-grmhd}. We consider a regular base grid of $79^3$ cells and allow up to two levels of adaptive mesh refinement. As before all kernel level optimisations have been switched on.

This scaling test was run on up to $56$ nodes of SuperMUC Phase~2 consisting of two $2.2$ GHz $14$ core Intel Xeon E5-v3 processors each. The tests were run using both Intel's TBB and MPI for parallelisation. As a reference value we use a test run on $56$ cores and calculate the corresponding parallel efficiency compared to this baseline. Table \ref{table-strongscaling} contains the results of this strong scaling test. Here, efficiency measures the speedup obtained divided by the expected optimal speedup. The efficiency is over $1.0$ in some cases can be explained by the chosen baseline of $56$ cores.

\begin{table}[h!]
\centering
\caption{Hybrid scaling of the GRMHD application. A $N=6$ ADER-DG approximation
  plus Finite Volumes limiting is used. Up to two levels of adaptive mesh refinement are employed.  }
\label{table-strongscaling}
\begin{tabular}{cccccccc}
 \toprule
 & & \multicolumn{2}{c}{regular} & \multicolumn{2}{c}{ $\Delta l=1$} & \multicolumn{2}{c}{$\Delta l = 2$} \\
 \cmidrule(lr){1-2} \cmidrule(lr){3-4} \cmidrule(lr){5-6}\cmidrule(lr){7-8}
 \# cores (nodes) & Threads & time & efficiency  & time & efficiency & time & efficiency\\
 \cmidrule(lr){1-2}
 \cmidrule(lr){3-8}
  56 (2)  & 14 & 91.93 &    1.00 & 139.43& 1.00  & 395.90 & 1.00 \\
  112 (4) & 7  & 37.68 & 	1.22 & 57.03 & 1.29	 & 151.11 &	1.31 \\
  224 (8) & 4  & 30.88 & 	0.77 & 34.17 & 0.89  & 139.42 &	0.93 \\
  448 (16)& 2  & 17.67 &	0.43 & 26.66 & 0.65	 & 57.02 &	0.85 \\
  784 (28)& 1  & 13.88 &	0.32 & 22.89 & 0.44	 & 34.17 &	0.67 \\
 \bottomrule
\end{tabular}
\end{table}

\section{Conclusions and Future Work}

This paper introduces a software engine that allows users to
write higher-order ADER-DG codes for hyperbolic PDE systems with
both conservative and non-conservative terms and a-posteriori limiters. 
The engine, 
ExaHyPE, has been designed to work on a wide range of computer systems from
laptops to large high-performance compute clusters.
Its core vision is to provide an engine rather than a framework: 
Technical details both on the computational science and numerical side are hidden from the user.
A specification file plus very few routines realising the PDE terms are
typically the only customisation points modified by user codes.
Users can focus on the physics.
To achieve this high level of abstraction, writing codes in ExaHyPE requires the use of a pre-specified set of numerical methods.
The code thus clearly stands in the tradition of software packages such as
Clawpack \cite{Clawpack:2017},   as opposed to 
more generic, 
general-purpose software packages such as AMReX, deal.II or DUNE. 

There are three general directions for future work. 
First, the sketched application areas have to make an impact in their respective
domain.
This comprises application-specific performance engineering and studies of real-world setups and data. It notably also comprises the coupling of engine applications with other code building blocks. 
Examples for first work into this direction are \cite{duru:curvilinear1,duru:curvilinear2,tavelli:dim}. 
Second, we have to continue to investigate and to invest into the methods
under the hood of ExaHyPE. 
Examples for such new ingredients on our agenda are accelerator support, local
time stepping, and more dynamic load balancing and
autotuning \cite{Schreiber:18:Invasic,Charrier:17}.
Finally, we plan extensions of the core paradigm of the engine.
The development of ExaHyPE from scratch has been possible as we restricted
ourself to ADER-DG only.
In the future, we plan to elaborate to which degree we are able to add
particle-based (Particle-in-Cell methods or simple tracers) algorithms or
multigrid solvers (for elliptic subterms) on top of ExaHyPE. 
This will open new user communities to the engine.
First feasibility studies for this
\cite{Reps:15:Helmholtz,Weinzierl:16:PIC,Weinzierl:18:BoxMG} already exist.

\section{Acknowledgements and Funding}
This project has received funding from the European Union's Horizon 2020 research and innovation programme under grant agreement No 671698, \url{www.exahype.eu}.

The ExaHyPE team acknowledges additional support by the European Union’s Horizon 2020 research and innovation program (ChEESE, grant no. 823844),
A.-A.G. specifically thanks the ERC (starting grant number: 852992, TEAR), the German Research Foundation (DFG) (projects no. KA 2281/4-1, GA 2465/2-1, GA 2465/3-1), KONWIHR – the Bavarian Competence Network for Technical and Scientific High Performance Computing (project NewWave), BaCaTec (project no. A4), and KAUST-CRG (GAST, grant no. ORS-2016- CRG5-3027 and FRAGEN, grant no. ORS-2017-CRG6 3389.02) for additional support.

The authors gratefully acknowledge the 
support by the Leibniz Supercomputing Centre (\url{www.lrz.de}), which also provided the computing resources on SuperMUC (grant no.~pr48ma and grant no.~pr63qo) .

We would especially like to thank the many people who have made contributions to ExaHyPE, in particular our previous
team members Benjamin Hazelwood, Angelika Schwarz, Vasco Varduhn and Olindo Zanotti.


\section*{References}
\bibliographystyle{elsarticle-num}
\bibliography{exa}

\begin{thebibliography}{10}
\expandafter\ifx\csname url\endcsname\relax
  \def\url#1{\texttt{#1}}\fi
\expandafter\ifx\csname urlprefix\endcsname\relax\def\urlprefix{URL }\fi
\expandafter\ifx\csname href\endcsname\relax
  \def\href#1#2{#2} \def\path#1{#1}\fi

\bibitem{Duru:2017}
K.~Duru, L.~Rannabauer, A.-A. Gabriel, H.~Igel, A new discontinuous {Galerkin}
  spectral element method for elastic waves with physically motivated numerical
  fluxes, arXiv:1802.06380 (2017).

\bibitem{tavelli:dim}
M.~{Tavelli}, M.~{Dumbser}, D.~E. {Charrier}, L.~{Rannabauer}, T.~{Weinzierl},
  M.~{Bader}, A simple diffuse interface approach on adaptive {Cartesian} grids
  for the linear elastic wave equations with complex topography, Journal of
  Computational Physics 386 (2019) 158--189.

\bibitem{Bishop:2016}
N.~T. Bishop, L.~Rezzolla, Extraction of gravitational waves in numerical
  relativity, Living Reviews in Relativity 19~(1) (2016) 2.
\newblock \href {http://dx.doi.org/10.1007/s41114-016-0001-9}
  {\path{doi:10.1007/s41114-016-0001-9}}.

\bibitem{dumbser:2017}
M.~{Dumbser}, F.~{Guercilena}, S.~{K{\"o}ppel}, L.~{Rezzolla}, O.~{Zanotti},
  {Conformal and covariant Z4 formulation of the Einstein equations: Strongly
  hyperbolic first-order reduction and solution with discontinuous Galerkin
  schemes}, Phys. Rev. D 97~(8) (2018) 084053.
\newblock \href {http://arxiv.org/abs/1707.09910} {\path{arXiv:1707.09910}},
  \href {http://dx.doi.org/10.1103/PhysRevD.97.084053}
  {\path{doi:10.1103/PhysRevD.97.084053}}.

\bibitem{zanotti:2016}
O.~Zanotti, M.~Dumbser, Efficient conservative {ADER} schemes based on {WENO}
  reconstruction and space-time predictor in primitive variable, Computational
  Astrophysics and Cosmology 3.

\bibitem{fambri:2017}
F.~Fambri, M.~Dumbser, O.~Zanotti, Space-time adaptive {ADER-DG} schemes for
  dissipative flows: Compressible {Navier-Stokes} and resistive {MHD}
  equations, Computer Physics Communications 219.

\bibitem{reed:1973}
W.~Reed, T.~R.~Hill, Triangular mesh methods for the neutron transport
  equation.

\bibitem{shu:1}
B.~Cockburn, C.-W. Shu, {The {R}unge-{K}utta local projection $P^1$ -
  discontinuous-{G}alerkin finite element method for scalar conservation laws},
  ESAIM: Mathematical Modelling and Numerical Analysis 25~(3) (1991) 337--361.

\bibitem{shu:2}
B.~Cockburn, C.-W. Shu, {TVB} {R}unge-{K}utta local projection discontinuous
  {G}alerkin finite element method for conservation laws {II}: General
  framework, Mathematics of Computation 52~(186) (1989) 411--435.

\bibitem{shu:3}
B.~Cockburn, S.-Y. Lin, C.-W. Shu, {TVB} {R}unge-{K}utta local projection
  discontinuous {G}alerkin finite element method for conservation laws {III}:
  One-dimensional systems, Journal of Computational Physics 84~(1) (1989) 90 --
  113.

\bibitem{shu:4}
B.~{Cockburn}, S.~{Hou}, C.-W. {Shu}, {The {R}unge-{K}utta local projection
  discontinuous {G}alerkin finite element method for conservation laws. {IV}.
  The multidimensional case}, Mathematics of Computation 54 (1990) 545--581.

\bibitem{shu:5}
B.~Cockburn, C.-W. Shu, The {R}unge–{K}utta discontinuous {G}alerkin method
  for conservation laws {V}: Multidimensional systems, Journal of Computational
  Physics 141~(2) (1998) 199 -- 224.

\bibitem{Toro2001}
E.~F. Toro, R.~C. Millington, L.~A.~M. Nejad,
  \href{https://doi.org/10.1007/978-1-4615-0663-8_87}{Towards Very High Order
  Godunov Schemes}, Springer US, Boston, MA, 2001, pp. 907--940.
\newblock \href {http://dx.doi.org/10.1007/978-1-4615-0663-8_87}
  {\path{doi:10.1007/978-1-4615-0663-8_87}}.
\newline\urlprefix\url{https://doi.org/10.1007/978-1-4615-0663-8_87}

\bibitem{Titarev:2002}
V.~A. Titarev, E.~F. Toro, {ADER: Arbitrary High Order Godunov Approach},
  Journal of Scientific Computing 17~(1) (2002) 609--618.
\newblock \href {http://dx.doi.org/10.1023/A:1015126814947}
  {\path{doi:10.1023/A:1015126814947}}.

\bibitem{Gassner:2011:ExplicitOneStep}
G.~Gassner, M.~Dumbser, F.~Hindenlang, C.-D. Munz, Explicit one-step time
  discretizations for discontinuous {Galerkin} and finite volume schemes based
  on local predictors, Journal of Computational Physics 230~(11) (2011)
  4232--4247.
\newblock \href {http://dx.doi.org/10.1016/j.jcp.2010.10.024}
  {\path{doi:10.1016/j.jcp.2010.10.024}}.

\bibitem{Dumbser:2008}
M.~Dumbser, C.~Enaux, E.~F. Toro, Finite volume schemes of very high order of
  accuracy for stiff hyperbolic balance laws, J. Comput. Phys. 227~(8) (2008)
  3971--4001.

\bibitem{hartmann:2002}
R.~Hartmann, P.~Houston, Adaptive discontinuous {G}alerkin finite element
  methods for the compressible {Euler} equations, Journal of Computational
  Physics 183~(2) (2002) 508 -- 532.

\bibitem{persson:2006}
P.-O. Persson, J.~Peraire, Sub-cell shock capturing for discontinuous
  {G}alerkin methods, in: 44th AIAA Aerospace Sciences Meeting and Exhibit, p.
  112.

\bibitem{rezzolla:2011}
D.~Radice, L.~Rezzolla, Discontinuous {G}alerkin methods for
  general-relativistic hydrodynamics: Formulation and application to
  spherically symmetric spacetimes, Phys. Rev. D 84 (2011) 024010.

\bibitem{shu:WENO}
J.~Qiu, C.~Shu, {R}unge--{K}utta discontinuous {G}alerkin method using {WENO}
  limiters, SIAM Journal on Scientific Computing 26~(3) (2005) 907--929.

\bibitem{shu:HWENO}
J.~Qiu, C.-W. Shu, Hermite {WENO} schemes and their application as limiters for
  runge-kutta discontinuous {G}alerkin method: one-dimensional case, Journal of
  Computational Physics 193 (2003) 115--135.

\bibitem{loubere:1}
R.~Loubère, M.~Dumbser, S.~Diot, A new family of high order unstructured
  {MOOD} and {ADER} finite volume schemes for multidimensional systems of
  hyperbolic conservation laws, Communications in Computational Physics 16~(3)
  (2014) 718–763.
\newblock \href {http://dx.doi.org/10.4208/cicp.181113.140314a}
  {\path{doi:10.4208/cicp.181113.140314a}}.

\bibitem{loubere:2}
S.~Diot, R.~Loub{\`e}re, S.~Clain, The {MOOD} method in the three-dimensional
  case: Very-high-order finite volume method for hyperbolic systems,
  International Journal of Numerical Methods in Fluids 73 (2013) 362--392.

\bibitem{loubere:3}
S.~Diot, S.~Clain, R.~Loub{\`e}re, A high-order finite volume method for
  systems of conservation laws -- multi-dimensional optimal order detection
  ({MOOD}), Journal of Computational Physics 230~(10) (2011) 4028 -- 4050.

\bibitem{loubere:4}
M.~Dumbser, O.~Zanotti, R.~Loub\`{e}re, S.~Diot, {A Posteriori Subcell Limiting
  of the Discontinuous {Galerkin} Finite Element Method for Hyperbolic
  Conservation Laws}, J. Comput. Phys. 278~(C) (2013) 47--75.

\bibitem{Dumbser:14:Posteriori}
M.~Dumbser, O.~Zanotti, R.~Loub\`{e}re, S.~Diot, A posteriori subcell limiting
  of the discontinuous {Galerkin} finite element method for hyperbolic
  conservation laws, Journal of Computational Physics 278 (2014) 47--75.

\bibitem{Weinzierl:11:Peano}
T.~Weinzierl, M.~Mehl, {Peano} -- {A} {Traversal} and {Storage} {Scheme} for
  {Octree}-{Like} {Adaptive} {Cartesian} {Multiscale} {Grids}, SIAM J. Sci.
  Comput. 33~(5) (2011) 2732--2760.
\newblock \href {http://dx.doi.org/10.1137/100799071}
  {\path{doi:10.1137/100799071}}.

\bibitem{Weinzierl:2019}
T.~Weinzierl, The {P}eano software---parallel, automaton-based, dynamically
  adaptive grid traversals, ACM Transactions on Mathematical SoftwareIn press.

\bibitem{Zanotti:2015:SpaceTimeAMR}
O.~Zanotti, F.~Fambri, M.~Dumbser, A.~Hidalgo, Space–time adaptive {ADER}
  discontinuous {Galerkin} finite element schemes with a posteriori sub-cell
  finite volume limiting, Computers \& Fluids 118 (2015) 204--224.

\bibitem{DIM2D}
M.~Dumbser, A simple two-phase method for the simulation of complex free
  surface flows, Computer Methods in Applied Mechanics and Engineering 200
  (2011) 1204--1219.

\bibitem{DIM3D}
M.~Dumbser, {A Diffuse Interface Method for Complex Three-Dimensional Free
  Surface Flows}, Computer Methods in Applied Mechanics and Engineering 257
  (2013) 47--64.

\bibitem{jenkins}
{The Jenkins Project}, \url{https://jenkins.io/}.

\bibitem{duru:curvilinear1}
K.~Duru, A.~Gabriel, G.~Kreiss, On energy stable discontinuous {G}alerkin
  spectral element approximations of the perfeclty matched layer for the wave
  equation, Computer Methods in Applied Mechanics and Engineering 350 (2019)
  898--937.

\bibitem{duru:curvilinear2}
K.~Duru, L.~Rannabauer, O.-K.~A. Ling, A.-G. Gabriel, H.~Igel, M.~Bader, A
  stable discontinuous {G}alerkin method for linear elastodynamics in
  geometrically complex media using physics based numerical fluxes.

\bibitem{Krenz:19:ExaCloud}
L.~Krenz, L.~Rannabauer, M.~Bader, \href{https://arxiv.org/abs/1905.05524}{A
  high-order discontinuous {G}alerkin solver with dynamic adaptive mesh
  refinement to simulate cloud formation processes}, in: Parallel Processing
  and Applied Mathematics, 13th International Conference PPAM 2019, Vol. 12043,
  2020.
\newline\urlprefix\url{https://arxiv.org/abs/1905.05524}

\bibitem{koeppel:grmhd}
S.~{K{\"o}ppel}, {Towards an exascale code for GRMHD on dynamical spacetimes}
  1031 (2018) 012017.
\newblock \href {http://arxiv.org/abs/1711.08221} {\path{arXiv:1711.08221}},
  \href {http://dx.doi.org/10.1088/1742-6596/1031/1/012017}
  {\path{doi:10.1088/1742-6596/1031/1/012017}}.

\bibitem{fambri:grmhd}
F.~Fambri, M.~Dumbser, S.~Köppel, L.~Rezzolla, O.~Zanotti, {ADER}
  discontinuous {G}alerkin schemes for general-relativistic ideal
  magnetohydrodynamics, Monthly Notices of the Royal Astronomical Society
  477~(4) (2018) 4543--4564.
\newblock \href
  {http://arxiv.org/abs//oup/backfile/content_public/journal/mnras/477/4/10.1093_mnras_sty734/1/sty734.pdf}
  {\path{arXiv:/oup/backfile/content_public/journal/mnras/477/4/10.1093_mnras_sty734/1/sty734.pdf}},
  \href {http://dx.doi.org/10.1093/mnras/sty734}
  {\path{doi:10.1093/mnras/sty734}}.

\bibitem{rezzolla:2013}
L.~{Rezzolla}, O.~{Zanotti}, {Relativistic Hydrodynamics}, Oxford University
  Press, Oxford UK, 2013.

\bibitem{Weinzierl:18:Peano}
T.~Weinzierl, The {P}eano software---parallel, automaton-based, dynamically
  adaptive grid traversals, ACM Trans. Math. Softw.(2nd version under review).

\bibitem{abgrall:1994}
R.~Abgrall, On essentially non-oscillatory schemes on unstructured meshes:
  Analysis and implementation, Journal of Computational Physics 114~(1) (1994)
  45 -- 58.

\bibitem{dumbser:weno}
M.~Dumbser, A.~Hidalgo, O.~Zanotti, High order space–time adaptive
  {ADER-WENO} finite volume schemes for non-conservative hyperbolic systems,
  Computer Methods in Applied Mechanics and Engineering 268 (2014) 359 -- 387.

\bibitem{muscl}
V.~P. Kolgan, Application of the principle of minimizing the derivative to the
  construction of finite-difference schemes for computing discontinuous
  solutions of gas dynamics, Journal of Computational Physics 230 (2011)
  2384--2390.

\bibitem{leer:1997}
B.~van Leer, Towards the ultimate conservative difference scheme, Journal of
  Computational Physics 135~(2) (1997) 229 -- 248.
\newblock \href {http://dx.doi.org/https://doi.org/10.1006/jcph.1997.5704}
  {\path{doi:https://doi.org/10.1006/jcph.1997.5704}}.

\bibitem{ader1}
V.~Titarev, E.~Toro, {ADER:} arbitrary high order {G}odunov approach, J. Sci.
  Comput. 17 (2002) 609--618.

\bibitem{Zanotti:2015}
O.~Zanotti, F.~Fambri, M.~Dumbser, A.~Hidalgo, {Space–time adaptive ADER
  discontinuous {G}alerkin finite element schemes with a posteriori sub-cell
  finite volume limiting}, Computers \& Fluids 118 (2015) 204 -- 224.
\newblock \href
  {http://dx.doi.org/https://doi.org/10.1016/j.compfluid.2015.06.020}
  {\path{doi:https://doi.org/10.1016/j.compfluid.2015.06.020}}.

\bibitem{Heinecke:16:LIBXSMM}
A.~Heinecke, G.~Henry, M.~Hutchinson, H.~Pabst, {LIBXSMM}: Accelerating small
  matrix multiplications by runtime code generation, in: SC16: International
  Conference for High Performance Computing, Networking, Storage and
  Analysis(SC), 2016, pp. 981--991.
\newblock \href {http://dx.doi.org/10.1109/SC.2016.83}
  {\path{doi:10.1109/SC.2016.83}}.

\bibitem{rannabauer:swe:2018}
L.~Rannabauer, S.~Haas, D.~E. Charrier, T.~Weinzierl, Michael, Simulation of
  tsunamis with the exascale hyperbolic {PDE} engine {ExaHyPE}, in:
  Environmental Informatics: Techniques and Trends. Adjunct Proceedings of the
  32nd edition of the EnviroInfo., 2018.

\bibitem{Weinzierl:RunLengthEncoding}
W.~Eckhardt, T.~Weinzierl, A blocking strategy on multicore architectures for
  dynamically adaptive {PDE} solvers, in: Parallel Processing and Applied
  Mathematics, 8th International Conference, {PPAM} 2009, Wroclaw, Poland,
  September 13-16, 2009. Revised Selected Papers, Part {I}, 2009, pp. 567--575.
\newblock \href {http://dx.doi.org/10.1007/978-3-642-14390-8\_59}
  {\path{doi:10.1007/978-3-642-14390-8\_59}}.

\bibitem{Charrier:18:Enclave}
D.~Charrier, B.~Hazelwood, T.~Weinzierl, Enclave tasking for discontinuous
  galerkin methods on dynamically adaptive meshes, SIAM Journal on Scientific
  Computing.

\bibitem{Dumbser:2018}
M.~Dumbser, F.~Fambri, M.~Tavelli, M.~Bader, T.~Weinzierl,
  \href{https://doi.org/10.3390/axioms7030063}{Efficient implementation of
  {ADER} {D}iscontinuous {G}alerkin schemes for a scalable hyperbolic {PDE}
  engine}, Axioms 7~(3) (2018) 63.
\newline\urlprefix\url{https://doi.org/10.3390/axioms7030063}

\bibitem{BaerNunziato1986}
M.~R. Baer, J.~W. Nunziato, A two-phase mixture theory for the
  deflagration-to-detonation transition {(DDT)} in reactive granular materials,
  J. Multiphase Flow 12 (1986) 861--889.

\bibitem{Gaburro2018}
E.~Gaburro, M.~Castro, M.~Dumbser, {A well balanced diffuse interface method
  for complex nonhydrostatic free surface flows}, Computers and Fluids 175
  (2018) 180--198.

\bibitem{NACA}
J.~Moran, \href{https://books.google.de/books?id=4eVP3yWZ1LgC}{An Introduction
  to Theoretical and Computational Aerodynamics}, Dover Books on Aeronautical
  Engineering, Dover Publications, 2003.
\newline\urlprefix\url{https://books.google.de/books?id=4eVP3yWZ1LgC}

\bibitem{Reinders:2007}
J.~Reinders, Intel Threading Building Blocks, 1st Edition, O'Reilly \&
  Associates, Inc., Sebastopol, CA, USA, 2007.

\bibitem{CharrierUndEkaterinasPaper2019}
D.~E. Charrier, B.~Hazelwood, E.~Tutlyaeva, M.~Bader, M.~Dumbser,
  A.~Kudryavtsev, A.~Moskovsky, T.~Weinzierl, Studies on the energy and deep
  memory behaviour of a cache-oblivious, task-based hyperbolic {PDE} solver,
  The International Journal of High Performance Computing Applications 33~(5)
  (2019) 973--986.
\newblock \href {http://dx.doi.org/10.1177/1094342019842645}
  {\path{doi:10.1177/1094342019842645}}.

\bibitem{Wolherr}
S.~Wollherr, A.-A. Gabriel, P.~M. Mai,
  \href{https://agupubs.onlinelibrary.wiley.com/doi/abs/10.1029/2018JB016355}{Landers
  1992 “reloaded”: Integrative dynamic earthquake rupture modeling},
  Journal of Geophysical Research: Solid Earth 124~(7)  6666--6702.
\newblock \href {http://dx.doi.org/10.1029/2018JB016355}
  {\path{doi:10.1029/2018JB016355}}.
\newline\urlprefix\url{https://agupubs.onlinelibrary.wiley.com/doi/abs/10.1029/2018JB016355}

\bibitem{Ullrich}
T.~Ulrich, A.-A. Gabriel, J.-P. Ampuero, W.~Xu, Dynamic viability of the 2016
  mw 7.8 kaikōura earthquake cascade on weak crustal faults, Nature
  Communications 10(1213).
\newblock \href {http://dx.doi.org/doi:10.1038/s41467-019-09125-w.}
  {\path{doi:doi:10.1038/s41467-019-09125-w.}}

\bibitem{seissol}
C.~Uphoff, S.~Rettenberger, M.~Bader, E.~H. Madden, T.~Ulrich, S.~Wollherr,
  A.-A. Gabriel, \href{https://dl.acm.org/citation.cfm?id=3126948}{Extreme
  scale multi-physics simulations of the tsunamigenic 2004 sumatra megathrust
  earthquake}, in: SC '17: Proceedings of the International Conference for High
  Performance Computing, Networking, Storage and Analysis, ACM, 2017.
\newline\urlprefix\url{https://dl.acm.org/citation.cfm?id=3126948}

\bibitem{Day:LOH1}
S.~M. Day, C.~R. Bradley, Memory-efficient simulation of anelastic wave
  propagation., Bulletin of the Seismological Society of America 3  520–531.
\newblock \href {http://dx.doi.org/https://doi.org/10.1785/0120000103}
  {\path{doi:https://doi.org/10.1785/0120000103}}.

\bibitem{Dumbser:2016:HLLEM}
M.~Dumbser, D.~S. Balsara, A new efficient formulation of the {HLLEM} {Riemann}
  solver for general conservative and non-conservative hyperbolic systems, J.
  Comput. Phys. 304~(C) (2016) 275--319.
\newblock \href {http://dx.doi.org/10.1016/j.jcp.2015.10.014}
  {\path{doi:10.1016/j.jcp.2015.10.014}}.

\bibitem{Rannabauer:2018}
L.~Rannabauer, M.~Dumbser, M.~Bader, {ADER-DG} with a-posteriori finite-volume
  limiting to simulate tsunamis in a parallel adaptive mesh refinement
  framework, Computers \& Fluids\href
  {http://dx.doi.org/https://doi.org/10.1016/j.compfluid.2018.01.031}
  {\path{doi:https://doi.org/10.1016/j.compfluid.2018.01.031}}.

\bibitem{michel:1972}
F.~C. {Michel}, {Accretion of Matter by Condensed Objects} 15 (1972) 153--160.
\newblock \href {http://dx.doi.org/10.1007/BF00649949}
  {\path{doi:10.1007/BF00649949}}.

\bibitem{Gassner:2008:ViscousDG}
G.~Gassner, F.~L\"{o}rcher, C.-D. Munz, A discontinuous {Galerkin} scheme based
  on~a~space-time expansion {II}. viscous flow equations in~multi dimensions,
  Journal of Scientific Computing 34~(3) (2008) 260--286.
\newblock \href {http://dx.doi.org/10.1007/s10915-007-9169-1}
  {\path{doi:10.1007/s10915-007-9169-1}}.

\bibitem{Tavelli:2016:StaggeredDG}
M.~Tavelli, M.~Dumbser, A staggered space{\textendash}time discontinuous
  {Galerkin} method for the three-dimensional incompressible
  {Navier}{\textendash}{Stokes} equations on unstructured tetrahedral meshes,
  Journal of Computational Physics 319 (2016) 294--323.
\newblock \href {http://dx.doi.org/10.1016/j.jcp.2016.05.009}
  {\path{doi:10.1016/j.jcp.2016.05.009}}.

\bibitem{Muller:2010:Cumulus}
A.~M{\"u}ller, J.~Behrens, F.~X. Giraldo, V.~Wirth, An adaptive discontinuous
  {Galerkin} method for modeling cumulus clouds, in: Fifth European Conference
  on Computational Fluid Dynamics, ECCOMAS CFD, 2010.

\bibitem{Clawpack:2017}
{Clawpack Development Team}, \href{http://www.clawpack.org}{Clawpack software},
  version 5.4.0 (2017).
\newblock \href {http://dx.doi.org/10.5281/zenodo.262111}
  {\path{doi:10.5281/zenodo.262111}}.
\newline\urlprefix\url{http://www.clawpack.org}

\bibitem{Schreiber:18:Invasic}
M.~Schreiber, T.~Weinzierl, A case study for a new invasive extension of
  {Intel's Threading Building Blocks}, in: HiPEAC 2018---3rd COSH Workshop on
  Co-Scheduling of HPC Applications, 2018.

\bibitem{Charrier:17}
D.~E. Charrier, T.~Weinzierl, An experience report on (auto-)tuning of
  mesh-based {PDE} solvers on shared memory systems, in: Parallel Processing
  and Applied Mathematics - 12th International Conference, {PPAM} 2017, Lublin,
  Poland, September 10-13, 2017, 2017, pp. 3--13.
\newblock \href {http://dx.doi.org/10.1007/978-3-319-78054-2\_1}
  {\path{doi:10.1007/978-3-319-78054-2\_1}}.

\bibitem{Reps:15:Helmholtz}
B.~Reps, T.~Weinzierl, A complex additive geometric multigrid solver for the
  {H}elmholtz equations on spacetrees, ACM Transactions on Mathematical
  Software 44~(1) (2017) 2:1--2:36.

\bibitem{Weinzierl:16:PIC}
T.~Weinzierl, B.~Verleye, P.~Henri, D.~Roose, Two particle-in-grid realisations
  on spacetrees, Parallel Computing 52 (2016) 42--64.

\bibitem{Weinzierl:18:BoxMG}
M.~Weinzierl, T.~Weinzierl, Quasi-matrix-free hybrid multigrid on dynamically
  adaptive {Cartesian} grids, ACM Trans. Math. Softw.(in press).

\end{thebibliography}







\newpage

 \appendix
 
 \section{Dependencies and prerequisites}

ExaHyPE requires the following prerequisites:
\begin{itemize}
  \item For sequential simulations, only a C++ compiler is required.
The code uses only few C++14 features, but for many older versions 
enabling those features through \texttt{--std=c++0x} is required.

  \item Python 3

  \item ExaHyPE's default build environment uses GNU Make.
\end{itemize}

Further, ExaHyPE has the following optional dependencies:
\begin{itemize}
  \item ExaHyPE user code can also be written in Fortran. In this case, a Fortran compiler is needed.
  
    \item To exploit multi- or manycore computers, Intel's TBB 2017 is required. It is open source and works with GCC and
Intel compilers.

  \item  To run ExaHyPE on a distributed memory cluster, MPI is needed. ExaHyPE uses only very basic MPI routines 
  (use e.g. MPI 1.3).

  \item To use ExaHyPE's optimised compute kernels, Intel's \texttt{libxsmm}\footnote{Libxsmm is open source: \url{https://github.com/hfp/libxsmm}} 
and Python's module \texttt{Jinja2}\footnote{Jinja2 is open source: \url{http://jinja.pocoo.org/}} are required. A local installation script is made available.
\end{itemize}

\section{Obtaining ExaHyPE}
ExaHyPE is free software is hosted at \url{www.exahype.eu}. There a two different options to obtain ExaHyPE, the first is to download a complete snapshot of ExaHyPE, in this case a snapshot of Peano is included. The second option is to clone the repository, in this case Peano has to be added manually. The repository is available at \url{https://gitlab.lrz.de/exahype/ExaHyPE-Engine}.
The ExaHyPE guidebook contains documentation, detailed rationale and links to further resources and is available at \url{http://www.peano-framework.org/exahype/guidebook.pdf}.


\end{document}